\documentclass[draftclsnofoot,twoside,onecolumn,letter,12pt]{IEEEtran}
\usepackage{graphicx,cite,amssymb,amsmath,psfrag,subfigure}
\usepackage{graphicx}
\usepackage{amssymb}
\usepackage{amsmath}
\usepackage[displaymath,mathlines]{lineno}
\usepackage{cite}
\usepackage{subfigure}
\usepackage{mathrsfs}
\usepackage{color}
\usepackage{tabulary}
\usepackage{multirow}

\newtheorem{theorem}{Theorem}
\newtheorem{lemma}{Lemma}
\newtheorem{corollary}{Corollary}
\newtheorem{proposition}{Proposition}

\newtheorem{remark}{Remark}

\DeclareMathOperator{\E}{\mathbb{E}}

\newcommand{\EXs}[2]{\E_{{#1}}\left\{{#2}\right\}}
\newcommand{\PDF}[2]{p_{{#1}}\left({#2}\right)}
\newcommand{\CG}[2]{\mathcal{CN}\left({#1},{#2}\right)}

\newcommand{\B}[1]{{\pmb{#1}}}

\newcommand{\Pu}{p_{\mathrm{u}}}

\newcommand{\Eu}{E_{\mathrm{u}}}

\begin{document}

%\vspace{-3 cm}
%\linenumbers
\title{
    \hspace{4cm}\\[-0.7cm]
    Uplink Performance Analysis of Multicell MU-MIMO Systems with ZF Receivers}
\author{
        Hien Quoc Ngo, Michail~Matthaiou, Trung Q. Duong, and
        Erik G. Larsson
\thanks{
        H.~Q.\ Ngo and E.~G.\ Larsson are with the Department of Electrical
    Engineering (ISY), Link\"{o}ping University, 581 83 Link\"{o}ping,
    Sweden
        (email: nqhien@isy.liu.se; egl@isy.liu.se).
}
\thanks{
M. Matthaiou is with the Department of Signals and Systems,
Chalmers University of Technology, 412 96, Gothenburg, Sweden
(email: michail.matthaiou@chalmers.se). }

\thanks{T. Q. Duong is with the Blekinge Institute of Technology 371
79, Karlskrona, Sweden (email: quang.trung.duong@bth.se). }

 \thanks{The work of H.~Q.\ Ngo and E.~G.\ Larsson was supported  in part by the Swedish Research Council
(VR), the Swedish
    Foundation for Strategic Research (SSF), and ELLIIT. E.~G.\ Larsson is a Royal
    Swedish Academy of Sciences (KVA) Research Fellow supported by a
    grant from the Knut and Alice Wallenberg Foundation. The work of M. Matthaiou was supported in part by the Swedish Governmental Agency for Innovation Systems
 (VINNOVA) within the VINN Excellence Center Chase.}

% \thanks{The work of M. Matthaiou has been supported in part by the Swedish Governmental Agency for Innovation Systems
% (VINNOVA) within the VINN Excellence Center Chase.}

\thanks{
        Parts of this work were presented at the 2011 IEEE Swedish Communication Technologies Workshop \cite{NDL:11:Swe-CTW}.
} }

\markboth{}
        {}

\maketitle

\renewcommand{\baselinestretch}{1.5} \normalsize

%%%%%%%%%%%%%%%%%%%%%%%%%%%%%%%%%%%%%%%%%%%%%%%%%%%%%%%%%%%%%%%%%%%%%
\vspace{-2cm}

\begin{abstract}
We consider the uplink of a multicell multiuser multiple-input
multiple-output system where the channel experiences both small
and large-scale fading. The data detection is done by using the
linear zero-forcing technique, assuming the base station (BS) has
perfect channel state information. We derive  new, exact
closed-form expressions for the uplink rate, symbol error rate,
and outage probability per user, as well as a lower bound on the
achievable rate. This bound is very tight and becomes exact in the
large-number-of-antennas limit. We further study the asymptotic
system performance in the regimes of high signal-to-noise ratio
(SNR), large number of antennas, and large number of users per
cell. We show that at high SNRs, the system is
interference-limited and hence, we cannot improve the system
performance by increasing the transmit power of each user.
Instead, by increasing the number of BS antennas, the effects of
interference and noise can be reduced, thereby improving the
system performance. We demonstrate that, with very large antenna
arrays at the BS, the transmit power of each user can be made
inversely proportional to the number of BS antennas while
maintaining a desired quality-of-service. Numerical results are
presented to verify our analysis.
\end{abstract}


\IEEEpeerreviewmaketitle

\section{Introduction}
\vspace{-0.1cm}
Multiple-input multiple-output (MIMO) technology can provide a
remarkable increase in data rate and reliability compared to
single-antenna systems. Recently, multiuser MIMO (MU-MIMO), where
the base stations (BSs) are equipped with multiple antennas and communicate with several
co-channel users, has gained much attention and is now being
introduced in several new generation wireless standards (e.g.,
LTE-Advanced, 802.16m) \cite{GKHCS:07:SPM}.

MU-MIMO systems have been studied from many perspectives including
communication, signalling, and information theory in both downlink
and uplink scenarios \cite{CM:WCOM:04,JG:IT:05}. %For example, the
%downlink capacity of a single user in a MU-MIMO system has been
%investigated for optimum receivers in \cite{McKay:IT:05}.
For the uplink, the maximum-likelihood multiuser detector can be
used to obtain optimal performance \cite{Ver:98:Book}. However,
this optimum receiver induces a significant complexity burden on
the system implementation, especially for large array
configurations. Therefore, linear receivers, in particular
Zero-Forcing (ZF) receivers, are of particular interest as
low-complexity alternatives
\cite{MCKN:11:COM,CW:07:JSAC,FMPHC:07:VT}. Note that all the above
mentioned works have only investigated a single-cell scenario,
where the effects of intercell interference have been neglected.
However, co-channel interference, appearing due to
frequency-reuse, represents an important impairment in cellular
systems. Recently, there has been an increasing research interest
in the performance of MU-MIMO in  interference-limited multi-cell
environments
\cite{CDG:00:COML,Blum:03:JSAC,DMP:04:WCOM,GHHSSY:10:JSAC,JAMV:11:WCOM}.
In fact, it has been shown that the capacity of the MU-MIMO
downlink can be dramatically reduced due to intercell interference
\cite{CDG:00:COML}.

Many interference cancellation and mitigation techniques have been
proposed for multicell MU-MIMO systems, such as maximum likelihood
multiuser detection \cite{DMP:04:WCOM}, BS cooperation
\cite{CA:08:WCOM}, and interference alignment \cite{SHT:11:COM}.
These techniques, however, have a high implementation complexity.
Very recently, there has been a great deal of interest in
multicell MU-MIMO, where the BSs are equipped with very large
antenna arrays
\cite{RPLLMET:11:SPM,Mar:10:WCOM,CD:10:SPM,NLM:11:TCOM,HBD:11:ACCCC}.
With very large antenna arrays, the intercell interference can be
successfully handled by using simple linear detectors, because the
channel vectors are nearly orthogonal when the number of antennas
is large (see e.g., \cite{RPLLMET:11:SPM,NLM:11:TCOM} for a more
detailed discussion). In this context, the asymptotic
signal-to-interference-plus-noise ratios (SINRs), when the number
of BS antennas grows infinite, were derived in \cite{Mar:10:WCOM}
for maximum-ratio combining (MRC) in the uplink and maximum-ratio
transmission in the downlink. In \cite{HBD:11:ACCCC}, by using
tools of random matrix theory, the authors derived a deterministic
approximation of the SINR for the uplink with MRC and minimum
mean-square error (MMSE) receivers, assuming that the number of
transmit antennas and number of users go to infinity at the same
rate. They also showed that the deterministic approximation of the
SINR is tight even with a moderate number of BS antennas and
users. However, since the limiting SINR obtained therein is
deterministic, this approximation does not enable us to further
analyze other figures of merit, such as the outage probability or
symbol error rate (SER). More importantly, iterative algorithms
are needed to compute the
 deterministic equivalent results. In \cite{NLM:11:TCOM},
lower bounds on the uplink achievable rates with linear detectors
were computed, and the authors showed that MRC performs as well as
ZF in a regime where the spectral efficiency is of the order of
$1$ bpcu per user. Nevertheless, it was demonstrated that  ZF
performs much better than MRC at higher spectral efficiencies.

%To the best of the authors' knowledge, there is no exactly
%analytical framework for analyzing the performance of uplink
%MU-MIMO systems in an interference-limited environment with a
%finite number of antennas.

Inspired by  the above discussion, we analyze in this paper the
performance of multicell MU-MIMO systems where many users
simultaneously transmit  data to a BS. The BS uses ZF to detect
the transmitted signals. Note that the MMSE receiver always
performs better than the ZF receiver. However, herein
 we consider ZF receivers for the following reasons: i) an exact analysis of MMSE receivers is a challenging mathematical problem in a multicell MU-MIMO setup. This implication can be seen by invoking the generic results of \cite{MCT:IT:10}; ii) the implementation of MMSE receivers requires additional
knowledge of the noise and interference statistics; iii) it is
well-known that ZF receivers perform equivalently to MMSE
receivers at high SNIRs \cite{TV:05:Book}; and iv) the performance
of ZF bounds that of MMSE from below, so the results we obtain
represent achievable lower bounds on the MMSE receivers'
performance. The paper makes the following specific contributions:
\begin{itemize}
\item We derive exact closed-form expressions for the ergodic data
rate, SER, and outage probability of the
uplink channel for any finite number of BS antennas. %in the forms of the exponential integral, confluent
%hypergeometric, and generalized hypergeometric functions which can
%be easily evaluated in most standard software packages, such as
%Matlab, Mathematica.
We also derive a tractable lower bound on the achievable rate.
Note that, although these exact results involve complicated
functions, they can be much more efficiently evaluated compared to
brute-force Monte-Carlo simulations.

\item Next, we focus on the ZF receiver's asymptotic performance
when the BS deploys a large antenna array. These results enable us
to explicitly study the effects of transmit power, intercell
interference, and number of BS antennas. For instance, when the
number of users per cell is fixed and the number of BS antennas
grows without bound, intercell interference and noise are averaged
out. However, when fixing the ratio between the number of BS
antennas and the number of users, the intercell interference does
not vanish when the number of antennas grows large. Yet, in both
cases by using very large antenna arrays, the transmit power of
each user can be made inversely proportional to the number of
antennas with no performance degradation.
\end{itemize}

\textit{Notation:} The superscript $H$ stands for conjugate
transpose. $\left[\B{A}\right]_{ij}$ denotes the ($i,j$)th entry
of a matrix $\B{A}$, and $\B{I}_{n}$ is the $n \times n$ identity
matrix. The expectation operation and the Euclidean norm are
denoted by $\mathbb{E}\left\{\cdot\right\}$ and $\| \cdot \|$,
respectively. The notation $\mathop \to \limits^{\tt a.s.}$ means
almost sure convergence. We use $a \mathop = \limits^{\tt d} b$ to
imply that $a$ and $b$ have the same distribution. Finally, we use
$\B{z} \sim \CG{\B{0}}{\B{\Sigma}}$ to denote a circularly
symmetric complex Gaussian vector $\B{z}$ with zero-mean and
covariance matrix $\B{\Sigma}$.

\section{Multicell MU-MIMO System} \label{sec: system}

In the following, we consider a multicell MU-MIMO system with $L$
cells. Each cell includes one BS equipped with $N$ antennas, and
$K$ single-antenna users ($N \geq K$). We consider uplink
transmission, and assume that the $L$ BSs share the same frequency
band. Conventionally, the communication between the BS and the
users is performed in separate time-frequency resources. However,
this approach inherently reduces the spectral efficiency and it is
therefore more efficient if several users communicate with the BS
in the same time-frequency resource
\cite{JAMV:11:WCOM,Mar:10:WCOM}. We assume that all users
simultaneously transmit data streams to their BSs. Therefore, the
$N \times 1$ received vector at the $l$th BS is given by
\begin{align} \label{eq MU-MIMO 1}
    \B{y}_l
    =
        \sqrt{p_{\mathrm{u}}}
        \sum_{i=1}^{L}
            \B{G}_{li}
            \B{x}_i
        +
        \B{n}_l
\end{align}
where $\B{G}_{li} \in \mathbb{C}^{N\times K}$ is the channel
matrix between the $l$th BS and the $K$ users in the $i$th cell,
i.e., $g_{limk} \triangleq \left[\B{G}_{li}\right]_{mk}$ is the
channel coefficient between the $m$th antenna of the $l$th BS and
the $k$th user in the $i$th cell; $\sqrt{p_{\mathrm{u}}} \B{x}_i
\in \mathbb{C}^{K\times 1}$ is the transmitted vector of $K$ users
in the $i$th cell (the average power transmitted by each user is
$p_{\mathrm{u}}$);  and $\B{n}_l \in \mathbb{C}^{N\times 1}$ is an
 additive white Gaussian noise (AWGN) vector, such that
$\B{n}_l \sim \CG{\B{0}}{\B{I}_M}$.  Note that, since the noise power
is assumed to be $1$, $\Pu$ can be considered as the normalized
``transmit" SNR and hence, it is
dimensionless.%\footnote{
%    $\B{z} \sim
%    \CG{0}{\B{\Sigma}}$ denotes a circularly symmetric complex
%    Gaussian vector with covariance matrix $\B{\Sigma}$ and
%    zero mean.
%    }

The channel matrix, $\B{G}_{li}$, models independent fast fading,
path-loss attenuation, and log-normal shadow fading. The
assumption of independent fast fading is sufficiently realistic
for systems where the antennas are sufficiently well separated
\cite{SA:00:Book}. Hence, its elements $g_{limk}$ are given by
\begin{align}\label{eq CM 1}
    g_{limk}
    =
        h_{limk} \sqrt{\beta_{lik}},~~~ m=1, 2, ..., N
\end{align}
where $h_{limk}$ is the fast fading coefficient from the $k$th
user in the $i$th cell to the $m$th antenna of the $l$th BS. The
coefficient $h_{limk}$ is assumed to be complex Gaussian
distributed with zero-mean and unit variance. Moreover,
$\sqrt{\beta_{lik}}$ represents the path-loss attenuation and
shadow fading which are assumed to be independent over $m$ and to
be constant over  many coherent intervals. This assumption is
reasonable since the distance between users and the BS is much
greater than the distance between the BS antennas. Additionally,
the validity of this assumption has been demonstrated in practice
even for large antenna arrays \cite{GERT:11:VTC}.

We assume that the BS has perfect channel state information (CSI).
This assumption is reasonable in an environment with low or
moderate mobility, so that long training intervals can be
afforded. Moreover, the results obtained under this assumption
serve as bounds on the performance for the case that the CSI is
imperfect due to estimation errors or feedback delays. We further
assume that the transmitted signals from the $K$ users in the
$l$th cell are detected using a ZF receiver. As such, the received
vector $\B{y}_l$ is processed by multiplying it with the
pseudo-inverse of $\B{G}_{ll}$ as:
\begin{align} \label{eq MU-MIMO 3}
    \B{r}_l
    &=
        \B{G}_{ll}^{\dagger}
        \B{y}_l
%    \nonumber
%    \\
    =
        \sqrt{\Pu}
        \B{x}_l
        +
        \sqrt{\Pu}
        \sum_{i \neq l}^{L}
            \B{G}_{ll}^{\dagger}
            \B{G}_{li}
            \B{x}_i
        + \B{G}_{ll}^{\dagger} \B{n}_l
\end{align}
where $\B{G}_{ll}^\dagger \triangleq \left(\B{G}_{ll}^H \B{G}_{ll}
\right)^{-1} \B{G}_{ll}^H$. Therefore, the $k$th element of
$\B{r}_l$ is given by
\begin{align} \label{eq MU-MIMO 3}
    \B{r}_{l,k}
    &=
        \sqrt{\Pu}
        \B{x}_{l,k}
        +
        \sqrt{\Pu}
        \sum_{i \neq l}^{L}
            \left[\B{G}_{ll}^{\dagger}\right]_k
            \B{G}_{li}
            \B{x}_i
        + \left[\B{G}_{ll}^{\dagger}\right]_k \B{n}_l
\end{align}
where $\B{x}_{l,k}$ is the $k$th element of $\B{x}_l$, which is
the transmitted signal from the $k$th user in the $l$th cell,
while $\left[\B{A}\right]_k$ denotes the $k$th row of a matrix
$\B{A}$. From \eqref{eq MU-MIMO 3}, the SINR of the uplink
transmission from the $k$th user in the $l$th cell to its BS is
defined as
\begin{align} \label{eq MU-MIMO 4}
    \gamma_k
    &\triangleq
        \frac{
            \Pu
            }{
            \Pu
            \sum_{i \neq l}^{L}
                \left\|
                    \left[\B{G}_{ll}^{\dagger}\right]_k
                    \B{G}_{li}
                \right\|^2
            +
                \left\|
                    \left[\B{G}_{ll}^{\dagger}\right]_k
                \right\|^2
            }.
\end{align}

\begin{proposition}\label{Prop 1}
The SINR of the uplink transmission from the $k$th user in the
$l$th cell to its BS can be represented as
\begin{align} \label{eq Prop1 1}
    \gamma_k
    \mathop = \limits^{\tt d}
                \frac{
                    \Pu X_k
                    }{
                    \Pu Z_l + 1
                    }
\end{align}
where $X_k$ and $Z_l$ are independent random
variables (RVs) whose probability density functions (PDFs) are respectively given by
\begin{align} \label{eq PDF1 1}
    \PDF{X_k}{x}
    &=
        \frac{
            e^{-x/\beta_{llk}}
            }{
            \left(N-K\right)!
            \beta_{llk}
            }
        \left(
            \frac{
                x
                }{
                \beta_{llk}
                }
        \right)^{N-K}, ~ x \geq 0
%\end{align}
%%
%
%%
%\begin{align}
    \\
    \label{eq PDF1 2}
    \PDF{Z_l}{z}
    &=
        \sum_{m=1}^{\varrho\left( \mathbf{\mathcal{A}}_l\right)}
        \sum_{n=1}^{\tau_m \left( \mathbf{\mathcal{A}}_l\right)} \!
            \mathcal{X}_{m,n} \left( \mathbf{\mathcal{A}}_l\right)
            \frac{
                \mu_{l,m}^{-n}
            }{
                \left(
                    n-1
                \right)
                !
            }
            z^{n-1}
            e^{\frac{-z}{\mu_{l,m}}}
       ,
        \quad
        z \geq 0
\end{align}
%
%
%where $$\mathbf{\mathcal{A}}_l = \left(%
%\begin{array}{cccccc}
%  \B{D}_{l1} &  & 0 & 0 & 0 & 0 \\
%  0 & \ddots & 0 & 0 & 0 & 0 \\
%  0 & 0 & \B{D}_{l(l-1)} & 0 & 0 & 0 \\
%  0 & 0 & 0 & \B{D}_{l(l+1)} & 0 & 0 \\
%  0 & 0 & 0 & 0 & \ddots & 0 \\
%  0 & 0 & 0 & 0 & 0 & \B{D}_{lL} \\
%\end{array}%
%\right)$$,
where $\mathbf{\mathcal{A}}_l \in \mathbb{C}^{K\left(L-1
\right)\times K\left(L-1 \right)}$ is given by
 {$$\mathbf{\mathcal{A}}_l \triangleq \left[%
\begin{array}{cccccc}
  \B{D}_{l1} &  &  &  &  &  \\
   & \ddots &  &  & \B{0} &  \\
   &  & \B{D}_{l(l-1)} &  &  &  \\
   & & & \B{D}_{l(l+1)} &  &  \\
  & \B{0} &  &  & \ddots &  \\
   &  &  &  &  & \B{D}_{lL} \\
\end{array}%
\right]$$} and $\varrho\left( \mathbf{\mathcal{A}}_l\right)$ is the
number of distinct diagonal elements of $\mathbf{\mathcal{A}}_l$;
$\mu_{l,1}, \mu_{l,2}, ..., \mu_{l,\varrho\left(
\mathbf{\mathcal{A}}_l\right)}$ are the distinct diagonal elements
in decreasing order; $\tau_m \left( \mathbf{\mathcal{A}}_l\right)$
is the multiplicity of $\mu_{l,m}$; and $\mathcal{X}_{m,n} \left(
\mathbf{\mathcal{A}}_l\right)$ is the $\left(m,n\right)$th
characteristic coefficient of $\mathbf{\mathcal{A}}_l$ which is
defined in \cite[Definition~4]{SW:08:IT}.
\begin{proof}
Dividing the denominator and nominator of \eqref{eq MU-MIMO 4} by
$\left\| \left[\B{G}_{ll}^{\dagger}\right]_k \right\|^2$, we
obtain
\begin{align} \label{eq Rate 1}
    \gamma_k
    &=
        \frac{
            \Pu
            \left\| \left[\B{G}_{ll}^{\dagger}\right]_k
            \right\|^{-2}
            }{
            \Pu
            \sum_{i \neq l}^{L}
                \left\|
                    \B{Y}_i
                \right\|^2
            +
            1
            }
\end{align}
where $\B{Y}_i \triangleq
\frac{\left[\B{G}_{ll}^{\dagger}\right]_k
\B{G}_{li}}{\left\|\left[\B{G}_{ll}^{\dagger}\right]_k \right\|}$.
Since $\left\| \left[\B{G}_{ll}^{\dagger}\right]_k \right\|^2 =
\left[\left(\B{G}_{ll}^H \B{G}_{ll} \right)^{-1} \right]_{kk},$
$\left\| \left[\B{G}_{ll}^{\dagger}\right]_k \right\|^{-2}$  has
an Erlang distribution with shape parameter $N-K+1$ and scale
parameter $\beta_{llk}$ \cite{GHP:02:CL}. Then,
\begin{align} \label{eq PDF1 1a}
    \left\| \left[\B{G}_{ll}^{\dagger}\right]_k \right\|^{-2}
    \mathop = \limits^{\tt d}
    X_k.
\end{align}
Conditioned on $\left[\B{G}_{ll}^\dagger\right]_k$, $\B{Y}_i$ is a
zero-mean complex Gaussian vector with covariance matrix
$\B{D}_{li}$ which is independent of
$\left[\B{G}_{ll}^\dagger\right]_k$. Therefore, $\B{Y}_i \sim
\CG{\B{0}}{\B{D}_{li}}$, where $\B{D}_{li}$ is a $K \times K$
diagonal matrix whose elements are given by
$\left[\B{D}_{li}\right]_{kk} = \beta_{lik}$. Then, $\sum_{i \neq
l}^{L}
                \left\|
                    \B{Y}_i
                \right\|^2$ is the sum of $K\left(L-1\right)$
statistically independent but not necessarily identically
distributed exponential RVs. Thus, from
\cite[Theorem~2]{BSW:07:WCOM}, we have that
\begin{align} \label{eq PDF1 1b}
    \sum_{i \neq l}^{L}
                \left\|
                    \B{Y}_i
                \right\|^2
    \mathop = \limits^{\tt d}
    Z_l.
\end{align}
From \eqref{eq Rate 1}--\eqref{eq PDF1 1b}, we can obtain
\eqref{eq Prop1 1}.
\end{proof}
\end{proposition}

\section{Finite-$N$ Analysis} \label{Sec:Rate}

In this section, we present exact analytical expressions for the
ergodic uplink rate, SER, and outage probability of the system
described in Section~\ref{sec: system}. We underline the fact that
the following results hold for any arbitrary number of BS antennas
$N \geq K$.

\subsection{Uplink Rate Analysis}\label{Sec:Rate_CF}

From Proposition~\ref{Prop 1}, the uplink ergodic rate from the
$k$th user in the $l$th cell to its BS (in bits/s/Hz) is
given by%\footnote{
%    In \cite{BZGO:10:SP}, a closed-form expression of the achievable ergodic rate was derived from
%    a stochastic behavior of the SINR which is the quotient of an
%    exponential distribution and the sum of independent but not identically exponential r.v.'s whereas here, our SINR is a quotient of an Erlang
%    distribution and the sum of i.n.i.d. exponential r.v.'s.
%}
%
%
\begin{align} \label{Rate 1}
    &\left\langle R_k \right\rangle
    =
        \EXs{X_k, Z_l}{
            \log_2
            \left(
                1+
                \frac{
                    \Pu X_k
                    }{
                    \Pu Z_l + 1
                }
            \right)
        }
    =
    \int_0^{\infty}
    \int_0^{\infty}
            \ln
            \left(
                1+
                \frac{
                    \Pu x
                    }{
                    \Pu z + 1
                }
            \right)
            \PDF{X_k}{x}
            \PDF{Z_l}{z}
    dx dz
    \nonumber
    \\
    &=
        \sum_{m=1}^{\varrho\left( \mathbf{\mathcal{A}}_l\right)}
        \sum_{n=1}^{\tau_m \left( \mathbf{\mathcal{A}}_l\right)}\!
            \frac{
                \mathcal{X}_{m,n} \left( \mathbf{\mathcal{A}}_l\right)
                \mu_{l,m}^{-n} \log_2 e
            }{
                \left(
                    n\!-\!1
                \right)
                !
                \left(N\!-\!K\right)!
                \beta_{llk}^{N-K+1}
            }
%        \nonumber
%        \\
%        & \hspace{-0.5cm} \times
    \!\int_0^{\infty}\!
    \int_0^{\infty}\!
            \ln
            \left(\!
                1\!+\!
                \frac{
                    \Pu x
                    }{
                    \Pu z \!+ \!1
                }
            \!\right)\!
            x^{N\!-\!K} e^{\frac{-x}{\beta_{llk}}}
            z^{n-1}
            e^{\frac{-z}{\mu_{l,m}}}
    dx dz.
\end{align}

We first evaluate the integral over $x$. By using
\cite[Eq.~(4.337.5)]{GR:07:Book}, we obtain
\begin{align} \label{Rate 2}
    \left\langle R_k \right\rangle
    &=
        \sum_{m=1}^{\varrho\left( \mathbf{\mathcal{A}}_l\right)}
        \sum_{n=1}^{\tau_m \left( \mathbf{\mathcal{A}}_l\right)}
        \sum_{p=0}^{N-K}
            \frac{
                \mathcal{X}_{m,n} \left( \mathbf{\mathcal{A}}_l\right)
                \mu_{l,m}^{-n} \log_2 e
            }{
                \left(
                    n\!-\!1
                \right)
                !
                \left(N\!-\!K\!-\!p\right)!
            }
        \int_0^{\infty}
        \left[
                \frac{
                    -
                    \left(\Pu z + 1 \right)^{N-K-p}
                    }{
                    \left(-\beta_{llk} \Pu\right)^{N-K-p}
                    }
                e^{\frac{\Pu z +1}{\beta_{llk} \Pu}}
                \mathrm{Ei}\left(\!-\frac{\Pu z +1}{\beta_{llk} \Pu} \right)
        \right.
        \nonumber
        \\
        & \hspace{3.5cm}
            \left.
                +
                \sum_{q=1}^{N-K-p}
                \left(q-1 \right)!
                \left(-\frac{\Pu z +1}{\beta_{llk} \Pu}\right)^{N-K-p-q}
            \right]
            z^{n-1}
            e^{-z/\mu_{l,m}}
    dz
\end{align}
 where
$\mathrm{Ei}\left(\cdot\right)$ is the exponential integral
function \cite[Eq.~(8.211.1)]{GR:07:Book}.

\begin{theorem}\label{Prop 2}
The uplink ergodic rate from the $k$th user in the $l$th cell to
its BS is given by
\begin{align} \label{eq Prop2 1}
    \left\langle R_k \right\rangle
    &\!=\!
        \log_2 e\!\!\!
        \sum_{m=1}^{\varrho\left( \mathbf{\mathcal{A}}_l\right)}
        \sum_{n=1}^{\tau_m \left( \mathbf{\mathcal{A}}_l\right)}\!
        \sum_{p=0}^{N\!-\!K}\!
            \frac{
                \mathcal{X}_{m,n}\!\! \left( \mathbf{\mathcal{A}}_l\!\right)
                \mu_{l,m}^{-n}
                \left(\!-1\!\right)^{N\!-K-p}
            }{
                \left(
                    n\!-\!1
                \right)
                !
                \left(N\!-\!K\!-\!p\right)!
            }
        \!\!\left[
            - e^{\frac{1}{\beta_{llk}\Pu}}
            \mathcal{I}_{n\!-\!1, N\!-\!K\!-\!p}\left(\!\frac{1}{\beta_{llk}}, \frac{1}{\beta_{llk}\Pu},\frac{1}{\mu_{l,m}}\!-\!\frac{1}{\beta_{llk}}\! \right)
        \right.
        \nonumber
        \\
        & \left.
                +
                \sum_{q=1}^{N-K-p}
                \frac{\left(q-1 \right)! \left(-1\right)^{q} \Pu^{-n}}
                    {\left(\beta_{llk} \Pu\right)^{N-K-p-q}}
                \Gamma\left(n\right)
                U\left(n, n+N+1-K-p-q, \frac{1}{\mu_{l,m}\Pu}\right)
            \right]
\end{align}
where $U\left(\cdot,\cdot,\cdot \right)$ is the confluent
hypergeometric function of the second kind
\cite[Eq.~(9.210.2)]{GR:07:Book},
\begin{align} \label{I Func}
    \mathcal{I}_{m,n}\left( a,b,\alpha \right)
    &\triangleq
        \sum_{i=0}^{m}
            \binom{m}{i}
            \left(-b\right)^{m-i}
        \left[
            \sum_{q=0}^{n+i}
            \frac{
                \left(n+i \right)^q
                b^{n+i-q}
                }{
                \alpha^{q+1} a^{m-q}
                }
            \mathrm{Ei} \left(\!-b\right)
            -
            \frac{
                \left(n+i \right)^{n+i}
                e^{\alpha b/a}
                }{
                \alpha^{n+i+1}
                a^{m-n-i}
                }
            \mathrm{Ei} \left(\!-\frac{\alpha b}{a}-b \!\right)
        \right.
    \nonumber
    \\
    & \hspace{3.5 cm}
        \left.
        +
        \frac{e^{-b}}{\alpha }
        \sum_{q=0}^{n+i-1}
        \sum_{j=0}^{n+i-q-1}
        \frac{
            j!
            \left(n+i\right)^q \binom{n+i-q-1}{j}
            b^{n+i-q-j-1}
            }{
            \alpha^q
            a^{m-q}
            \left(\alpha/a + 1 \right)^{j+1}
            }
        \right].
\end{align}
%
%and
%%
%\begin{align} \label{F Func} \setcounter{equation}{12}
%    \mathcal{F}_{m,n}\left( a,\alpha \right)
%    =
%        \sum_{i=0}^{n}
%        \binom{n}{i}
%        \left(m+i\right)!
%        \frac{a^i}{\alpha^{m+i+1}}.
%\end{align}
%
\begin{proof}
See Appendix~\ref{sec:proof:prop rate1}.
\end{proof}
\end{theorem}

In practice, users are located randomly within cells, such that
the large-scale fading factors for different users are different.
This results in all diagonal elements of $\mathbf{\mathcal{A}}_l$
being distinct. The following corollary corresponds to this
practically important special case.

%The following corollaries are two extreme cases of
%Theorem~\ref{Prop 2}.

\begin{corollary}\label{coro1}
If all diagonal elements of $\mathbf{\mathcal{A}}_l$ are
distinct, the ergodic rate in \eqref{eq Prop2 1} reduces to
\begin{align} \label{eq coro1 1}
    \left\langle R_k \right\rangle
    &\!=\!
        \log_2 e \!\!
        \sum_{m=1}^{K \left(L\!-\!1\right)}
        \sum_{p=0}^{N\!-\!K}
            \frac{
    \prod_{n=1,n\neq m}^{K\left(L-1\right)}
    \left(
        1
        -
        \mu_{l,n}/\mu_{l,m}
    \right)^{-1}
            }{
               \left(N\!-\!K\!-\!p\right)!
               \left(\!-1\!\right)^{N\!-K-p}
               \mu_{l,m}\
            }\!
        \left[
            - e^{\frac{1}{\beta_{llk}\Pu}}
            \mathcal{I}_{0, N\!-\!K\!-\!p}\left(\!\frac{1}{\beta_{llk}}, \frac{1}{\beta_{llk}\Pu},\frac{1}{\mu_{l,m}}\!-\!\frac{1}{\beta_{llk}}\! \right)
        \right.
        \nonumber
        \\
        &
            \left.
                +
                \sum_{q=1}^{N-K-p}
                \frac{\left(q-1 \right)! \left(-1\right)^{q} }
                    {\beta_{llk}^{N-K-p-q}}
                e^{\frac{1}{\mu_{l,m}\Pu}}
                \mu_{l,m}^{N+1-K-p-q}
                \Gamma\left(N+1-K-p-q, \frac{1}{\mu_{l,m}\Pu} \right)
                %U\left(1, 2+M-K-p-q, \frac{1}{\mu_{l,m}\Pu}\right)
            \right]
\end{align}
where $\Gamma(a,x)=\int_x^{\infty}t^{a-1}e^{-t}dt$ being the upper
incomplete gamma function \cite[Eq.~(8.350.2)]{GR:07:Book}.
\begin{proof}
For this case, substituting  $\varrho\left(
\mathbf{\mathcal{A}}_l\right) = K\left(L-1\right)$, $\tau_m \left(
\mathbf{\mathcal{A}}_l\right)=1$, and
\begin{align*}
    \mathcal{X}_{m,1} \left( \mathbf{\mathcal{A}}_l\right)
    =
    \prod_{n=1, n \neq m}^{K\left(L-1\right)}
    \left(
        1
        -
        \frac{\mu_{l,n}}{\mu_{l,m}}
    \right)^{-1}
\end{align*}
into \eqref{eq Prop2 1}, and using the identity
$U\left(1,a,x\right) = e^x x^{1-a}\Gamma\left(a-1,x\right)$
\cite[Eq. (07.33.03.0014.01)]{Wolfram}, we can obtain \eqref{eq
coro1 1}.
\end{proof}
\end{corollary}

In addition to the exact result given by Theorem~\ref{Prop 2}, we
now derive an analytical lower bound on the ergodic achievable
rate which is easier to evaluate:

\begin{proposition}\label{Prop BoundRate 1}
The uplink ergodic rate from the $k$th user in the $l$th cell to
its BS is lower bounded by
\begin{align} \label{eq PropBound 1}
    \left\langle R_k \right\rangle
    &\geq
    \log_2
    \Bigg(
        1+\Pu\beta_{llk}\exp
    \Bigg(
    \psi(N-K+1)
%    \right.
%    \right.
    \nonumber
    \\
    &
    \hspace{3cm}
    \left.
    \left.
    -
    \Pu\sum_{m=1}^{\varrho\left( \mathbf{\mathcal{A}}_l\right)}
\sum_{n=1}^{\tau_m \left( \mathbf{\mathcal{A}}_l\right)} \mu_{l,m}
n \mathcal{X}_{m,n} \left( \mathbf{\mathcal{A}}_l\right)
{}_3F_1\left(n+1,1,1;2;- \Pu\mu_{l,m}\right)
    \right)
    \right)
\end{align}
where $\psi(x)$ is Euler's digamma function
\cite[Eq.~(8.360.1)]{GR:07:Book}, and $_p{F}_q(\cdot)$ represents
the generalized hypergeometric function with $p,q$ non-negative
integers~\cite[Eq.~(9.14.1)]{GR:07:Book}.
\begin{proof}
See Appendix~\ref{sec:proof:propBound Rate}.
\end{proof}
\end{proposition}

\begin{remark}\label{remark 1}
From \eqref{eq Prop1 1}, we have that
\begin{align}\label{eq Asy 1}
    \lim_{\Pu \rightarrow \infty} \gamma_k
    \mathop = \limits^{\tt d}
            \frac{
            X_k
            }{
            \sum_{i \neq l}^{L}
                \left\|
                    \B{Y}_i
                \right\|^2
            }.
\end{align}
The above result explicitly demonstrates that the SINR is bounded
when $\Pu$ goes to infinity. This means that at high SNRs, we
cannot improve the system performance by simply increasing the
transmitted power of each user. The reason is that, when $\Pu$
increases, both the desired signal power and the interference
power increase.
\end{remark}

\subsection{SER Analysis}
In this section, we analyze the SER performance of the uplink for
each user. Let $\mathcal{M}_{\gamma_k}\left(s\right)$ be the
moment generating function (MGF) of $\gamma_k$. Then, using the
well-known MGF-based approach \cite{SA:00:Book}, we can deduce the
exact average SER of $M$-ary phase-shift keying ($M$-PSK) as
follows:

\begin{theorem} \label{prop: SER}
The average SER of the uplink from the $k$th user in the $l$th
cell to its BS for $M$-PSK is given by
\begin{align}\label{eq SER 1}
    {\tt SER}_{k}
    =
        \frac{1}{\pi}
        \int_{0}^{\Theta}
        \mathcal{M}_{\gamma_k}
            \left(
            \frac{g_{\tt MPSK}}{\sin^2\theta}
            \right)
        d \theta
\end{align}
where $\Theta \triangleq \pi -\frac{\pi}{M}$, $g_{\tt MPSK}
\triangleq \sin^2 \left(\pi/M\right)$, and
\begin{align}\label{eq SER 2}
    \mathcal{M}_{\gamma_k}\!\left(\!s\!\right)
    &\!=\!\!
        \sum_{m=1}^{\varrho\left(\! \mathbf{\mathcal{A}}_l\!\right)}
        \sum_{n=1}^{\tau_m \left( \!\mathbf{\mathcal{A}}_l\!\right)}
        \sum_{p=0}^{N\!-\!K\!+\!1}\!\!\!
        \binom{N\!\!-\!\!K\!\!+\!\!1}{p}
            \mathcal{X}_{m,n}\! \left( \mathbf{\mathcal{A}}_l\right)\!\!
        \left(\frac{-\beta_{llk} s}{\beta_{llk}s\! + \! 1/\Pu} \right)^{p}
%    \nonumber
%    \\
%    &\times
        {}_2F_{0} \left(\! n, p; \text{---}; \frac{-\mu_{l,m}}{1/\Pu\! + \!\beta_{llk} s} \!
        \right).
\end{align}
\begin{proof}
See Appendix~\ref{sec:proof:prop SER}.
\end{proof}
\end{theorem}

It is also interesting to investigate the SER at high SNRs in
order to obtain the diversity gain of the system under
consideration. For this case ($\Pu \to \infty$), by ignoring
$1/\Pu$ in \eqref{eq SER 2}, we obtain the asymptotic SER at high
SNRs as
\begin{align}\label{eq SER 1a}
    {\tt SER}_{k}^{\infty}
    =
        \frac{1}{\pi}
        \int_{0}^{\Theta}
        \mathcal{M}_{\gamma_k}^{\infty}
            \left(
            \frac{g_{\tt MPSK}}{\sin^2\theta}
            \right)
        d \theta
\end{align}
where
\begin{align}\label{eq SER 2a}
    \mathcal{M}_{\gamma_k}^{\infty}\left(s\right)
    &=
        \sum_{m=1}^{\varrho\left( \mathbf{\mathcal{A}}_l\right)}
        \sum_{n=1}^{\tau_m \left( \mathbf{\mathcal{A}}_l\right)}
        \sum_{p=0}^{N-K+1}
        \binom{N\!-\!K\!+\!1}{p}
            \mathcal{X}_{m,n}\! \left( \mathbf{\mathcal{A}}_l\right)
        \left(-1 \right)^{p}
%    \nonumber
%    \\
%    &\times
        {}_2F_{0} \left(\! n, p; \text{---}; \frac{-\mu_{l,m}}{\!\beta_{llk} s} \!
        \right).
\end{align}
This implies that at high SNRs, the SER converges to a constant
value that is independent of SNR; hence the diversity order, which
is defined as $\lim_{\Pu \to \infty} \frac{-\log {\tt
SER}_{k}}{\log\left(\Pu\right)}$, is equal to zero. This
phenomenon occurs due to the presence of interference. The
following corollary corresponds to the interesting case when all
diagonal elements of $\mathbf{\mathcal{A}}_l$ are distinct.

\begin{corollary} \label{coro 3}
If all diagonal elements of $\mathbf{\mathcal{A}}_l$ are distinct,
the exact and high-SNR MGF expressions in \eqref{eq SER 2} and
\eqref{eq SER 2a} reduce respectively to
\begin{align}\label{eq coro3 MGF 1}
    \mathcal{M}_{\gamma_k}\!\left(s\right)
    &=
        \sum_{m=1}^{K\left(L-1\right)}
        \sum_{p=0}^{N\!-\!K\!+\!1}\!\!\!
        \binom{N\!\!-\!\!K\!\!+\!\!1}{p}
            \mathcal{X}_{m,1}\! \left( \mathbf{\mathcal{A}}_l\right)\!\!
        \frac{\left(-\beta_{llk} s\right)^{p} \mu_{l,m} ^{-1}}{\left(1/\Pu\! + \!\beta_{llk} s \right)^{p-1}}
        e^{\frac{1/\Pu\! + \!\beta_{llk} s }{\mu_{l,m}}}
        \!{\tt E}_p\left(\!\frac{1/\Pu\! + \!\beta_{llk} s }{\mu_{l,m}} \!\right)
    &\\ \label{eq coro3 MGF 2}
    \mathcal{M}_{\gamma_k}^{\infty}\left(s\right)
    &=
        \sum_{m=1}^{K\left(L-1\right)}
        \sum_{p=0}^{N\!-\!K\!+\!1}\!\!\!
        \binom{N\!\!-\!\!K\!\!+\!\!1}{p}
            \mathcal{X}_{m,1}\! \left( \mathbf{\mathcal{A}}_l\right)\!\!
        \frac{\left(-1\right)^{p}\beta_{llk} s}{\mu_{l,m}}
        e^{\frac{\!\beta_{llk} s }{\mu_{l,m}}}
        {\tt E}_p\left(\frac{\!\beta_{llk} s }{\mu_{l,m}} \!\right)
\end{align}
where ${\tt{E}}_{n}(z)=\int_1^{\infty} t^{-n}e^{-zt}dt,~n=0, 1, 2,
\ldots, \mathrm{{Re}}(z)>0$, is the exponential integral function
of order $n$ \cite[Eq. (06.34.02.0001.01)]{Wolfram}.
\begin{proof}
Following a similar methodology as in Corollary~\ref{coro1} and
using the identity
\begin{align}\label{eq coro3 1}
    {}_2F_{0} \left(1, p; \text{---}; -x \right)
    =
    %\left(\frac{1}{x}\right)^a e^{1/x} \Gamma\left(1-a,\frac{1}{x} \right)
    \frac{1}{x} e^{1/x} {\tt E}_p \left(\frac{1}{x} \right)
\end{align}
we arrive at the desired results \eqref{eq coro3 MGF 1} and
\eqref{eq coro3 MGF 2}. Note that \eqref{eq coro3 1} is obtained
by using \cite[Eq.~(8.4.51.1)]{PBM:90:Book:v3},
\cite[Eq.~(8.2.2.15)]{PBM:90:Book:v3},
\cite[Eq.~(8.4.16.14)]{PBM:90:Book:v3} and
\cite[Eq.~(46)]{SL:03:IT}.
\end{proof}
\end{corollary}

From \eqref{eq SER 1}, we can see that to compute the SER we have
to perform a finite integration over $\theta$. To avoid this
integration, we can apply the tight approximation of
\cite{MZCC:09:COM} on \eqref{eq SER 1}, to get
\begin{align}\label{eq SER 2c}
    {\tt SER}_{k}
    \approx
        \left(
            \frac{\Theta}{2\pi}
            -
            \frac{1}{6}
        \right)
        \mathcal{M}_{\gamma_k}
            \left(
            g_{\tt MPSK}
            \right)
        +
        \frac{1}{4}
        \mathcal{M}_{\gamma_k}
            \left(
            \frac{4 g_{\tt MPSK}}{3}
            \right)
        +
        \left(
            \frac{\Theta}{2\pi}
            -
            \frac{1}{4}
        \right)
        \mathcal{M}_{\gamma_k}
            \left(
            \frac{g_{\tt MPSK}}{\sin^2\Theta}
            \right).
\end{align}
Clearly, the above expression is easier to evaluate compared to \eqref{eq SER 1}.

\subsection{Outage Probability Analysis}
The main goal of this section is to analytically assess the outage
probability of multicell MU-MIMO systems with ZF processing at the
BS. Especially for the case of non-ergodic channels (e.g.
quasi-static or block-fading), it is appropriate to resort to the
notion of outage probability to characterize the system
performance. The outage probability, $P_{{\tt out}}$, is defined
as the probability that the instantaneous SINR, $\gamma_k$, falls
below a given threshold value $\gamma_{{\tt th}}$, i.e.,
\begin{align}\label{eq Out 1}
 P_{{\tt out}}
 \triangleq
    {\tt Pr}
    \left(
        \gamma_k\leq\gamma_{{\tt th}}
    \right).
\end{align}

With this definition in hand, we can present the following novel,
exact result:
\begin{theorem} \label{pro: out}
The outage probability of transmission from the $k$th user in the
$l$th cell to its BS is given by
\begin{align}\label{eq Pro Out 1}
    P_{{\tt out}}
        =
        1
        &-
        \exp
            \left(
                -\frac{\gamma_{{\tt th}}}{\Pu\beta_{llk}}
            \right)
        \nonumber
        \\
        &
        \times
        \sum_{m=1}^{\varrho\left( \mathbf{\mathcal{A}}_l\right)}
        \sum_{n=1}^{\tau_m \left( \mathbf{\mathcal{A}}_l\right)}
        \sum_{p=0}^{N-K}
        \sum_{q=0}^{p}
            \binom{p}{q}
            \frac{
                \left(\frac{\gamma_{{\tt th}}}{\beta_{llk}}\right)^p
                }{
                p!
                }
            \!\mathcal{X}_{m,n}
            \left(
                \mathbf{\mathcal{A}}_l
            \right)
            \frac{\mu_{l,m}^{-n}}{\left(n-1\right)!}
            \frac{
                    \Gamma\left(n+q \right) \Pu^{q-p}
                }{
                \left(1/\mu_{l,m} +  \gamma_{{\tt th}}/\beta_{llk}\right)^{n+q}
                }.
\end{align}
\begin{proof}
See Appendix~\ref{sec:proof:Prop out}.
\end{proof}
\end{theorem}

Note that the exponential integral function and confluent
hypergeometric functions appearing in Theorems~\ref{Prop 2},
\ref{prop: SER}, and \ref{pro: out}  are built in functions and
can be easily evaluated by standard mathematical software
packages, such as MATHEMATICA or MATLAB. We now recall that we are
typically interested in small outage probabilities (e.g., in the
order of 0.01, 0.001 etc). In this light, when $\gamma_{{\tt th}}
\to 0$, we can obtain the following asymptotic result:
\begin{align}\label{eq Pro Out 1b}
    P_{{\tt out}}^{\infty}
        &=
        1
        -
        \sum_{m=1}^{\varrho\left( \mathbf{\mathcal{A}}_l\right)}
        \sum_{n=1}^{\tau_m \left( \mathbf{\mathcal{A}}_l\right)}
        \sum_{p=0}^{N-K}
            \frac{
                \left(\frac{\gamma_{{\tt th}}}{\beta_{llk}}\right)^p
                }{
                p!
                }
            \!\mathcal{X}_{m,n}
            \left(
                \mathbf{\mathcal{A}}_l
            \right)
            \frac{\mu_{l,m}^{-n}}{\left(n-1\right)!}
            \frac{
                    \Gamma\left(n+p \right)
                }{
                \left(1/\mu_{l,m} +  \gamma_{{\tt th}}/\beta_{llk}\right)^{n+p}
                }.
\end{align}
The above result is obtained by keeping the dominant term $p=q$ in
\eqref{eq Pro Out 1} and  letting $\gamma_{{\tt th}} \to 0$.
Similarly to the SER case, $P_{{\tt out}}^{\infty}$ is independent
of the SNR, thereby reflecting the deleterious impact of
interference. Furthermore, for the case described in
Corollaries~\ref{coro1} and \ref{coro 3}, we can get the following
simplified results:
\begin{corollary} \label{coro 4}
If all diagonal elements of $\mathbf{\mathcal{A}}_l$ are distinct,
the exact and high-SNR outage probability expressions in \eqref{eq
Pro Out 1} and \eqref{eq Pro Out 1b} reduce respectively to
\begin{align}\label{eq coro4 OP 1}
    P_{{\tt out}}
        &=
        1
        -
        \exp
            \left(
                -\frac{\gamma_{{\tt th}}}{\Pu\beta_{llk}}
            \right)
        \sum_{m=1}^{K\left( L-1\right)}
        \sum_{p=0}^{N-K}
        \sum_{q=0}^{p}
            \frac{
                \left(\frac{\gamma_{{\tt th}}}{\beta_{llk}}\right)^p
                }{
                \left(p-q\right)!
                }
            \frac{\mathcal{X}_{m,1}
            \left(
                \mathbf{\mathcal{A}}_l
            \right)}{\mu_{l,m}}
            \frac{
                     \Pu^{q-p}
                }{
                \left(1/\mu_{l,m} +  \gamma_{{\tt th}}/\beta_{llk}\right)^{q+1}
                }
    &\\ \label{eq coro4 OP 2}
    P_{{\tt out}}^{\infty}
        &=
        1
        -
        \sum_{m=1}^{K\left( L-1\right)}
        \sum_{p=0}^{N-K}
                \left(\frac{\gamma_{{\tt th}}}{\beta_{llk}}\right)^p
            \frac{\mathcal{X}_{m,1}
            \left(
                \mathbf{\mathcal{A}}_l
            \right)}{\mu_{l,m}}
            \frac{
                    1
                }{
                \left(1/\mu_{l,m} +  \gamma_{{\tt th}}/\beta_{llk}\right)^{1+p}
                }.
\end{align}
\end{corollary}

\section{Asymptotic ($N\to \infty$) Analysis}

As discussed in Remark~\ref{remark 1}, we cannot improve the
multicell MU-MIMO system performance by simply increasing the transmit power.
However,   we can improve the system performance by using  a large number of
BS antennas. Due to the array gain and diversity effects,
when $N$ increases,
  the received powers of both  the desired and the interference signals
increase. Yet, based on the asymptotic orthogonal property of the
channel vectors between the users and the BS, when $N$ is large,
the interference can be significantly reduced even with a simple
ZF receiver \cite{NLM:11:TCOM,Mar:10:WCOM}. In this section, we
analyze the asymptotic performance for large $N$.
% We derive asymptotic SINR results for large $N$, which are directly used to deduce the
%asymptotic achievable rate, SER, etc, by using the continuous
%mapping theorem \cite[Theorem~2.3]{Vaa:00:Book}.
We assume that
when $N$ increases, the elements of the channel
    matrix are still independent. To guarantee the independence of the channels, the antennas have to be sufficiently well separated.
    Note that the physical size of the antenna array can be small even with very large
    $N$. For example, at $2.6$ GHz, a cylindrical array with $128$ antennas, which
    comprises
    $4$ circles of $16$ dual polarized antenna elements (distance between adjacent
antennas is about $6$ cm which is half a wavelength), occupies
only a physical size of $28$ cm $\times$ $29$ cm
\cite{GERT:11:VTC}.

\subsubsection{Fixed $\Pu$, $K$, and $N\rightarrow \infty$}

Intuitively, when the number of BS antennas $N$ grows large, the
random vectors between the BS and the users as well as the noise
vector at the BS become pairwisely orthogonal and hence, the
interference from other users can be cancelled out. At the same
time, due to the array gain effect, thermal noise cancels out too.
This intuition is confirmed by the following analysis.

Since $X_k$  has an Erlang distribution with shape parameter
$N-K+1$ and scale parameter $\beta_{llk}$, $X_k$ can be
represented as
\begin{align}\label{eq Asy 2}
    X_k
    =
        \frac{\beta_{llk}}{2}
        \sum_{i=1}^{2\left(N-K+1\right)}
        Z_i^2
\end{align}
where $Z_1, Z_2, ..., Z_{2\left(N-K+1\right)}$ are independent,
standard normal RVs. Substituting \eqref{eq Asy 2} into \eqref{eq
Rate 1}, and dividing the denominator and the numerator of
$\gamma_k$ by $2 \left(N-K+1\right)$, we obtain
\begin{align} \label{eq Asy 3}
    \gamma_k
    &=
        \frac{
            \Pu
            \frac{\beta_{llk}}{2}
            \sum_{i=1}^{2\left(N-K+1\right)}
            Z_i^2/\left(2\left(N-K+1\right) \right)
            }{
            \left(
            \Pu
            \sum_{i \neq l}^{L}
                \left\|
                    \B{Y}_i
                \right\|^2
            +
            1
            \right)/\left(2\left(N-K+1\right) \right)
            }
%    %\nonumber
%    \\ \label{eq Asy 3}
    \mathop \rightarrow \limits^{\tt a.s.}  \infty, ~ \text{as} ~ N \rightarrow \infty
\end{align}
where \eqref{eq Asy 3} is obtained by using the law of large
numbers, i.e., the numerator converges to $ \Pu\beta_{llk}/2$,
while the denominator converges to $0$. The above result reveals
that when the number of BS antennas goes to infinity, the effects
of interference and noise disappear. Therefore, by increasing $N$,
the SINR grows without limit. Similar conclusions were presented
in \cite{NLM:11:TCOM}.

\subsubsection{Fixed $\Pu$, $\kappa=N/K$, and $N\rightarrow \infty$}
This is an interesting asymptotic scenario since in practice, the
number of BS antennas, $N$, is large but may not be much greater
than the number of users $K$. For this case, the property stating
that the channel vectors between users and the BS are pairwisely
orthogonal when $N\to \infty$ is not valid. In other words,
$\B{H}_{li}^H\B{H}_{li}$ does not converge point-wisely to an
``infinite-size identity matrix" \cite{CD:10:SPM}. Therefore, the
intercell interference cannot be cancelled out. Since $\B{Y}_i
\sim \CG{\B{0}}{\B{D}_{li}}$, it can be represented as
\begin{align} \label{eq Asy 3c}
    \B{Y}_i
    =
    \B{w}_i^H \B{D}_{li}^{1/2}
\end{align}
where $\B{w}_i \sim \CG{\B{0}}{\B{I}_{K}}$. From \eqref{eq Rate 1},
\eqref{eq Asy 2}, and \eqref{eq Asy 3c}, $\gamma_k$ can be
expressed as
\begin{align} \label{eq Asy 3d}
    \gamma_k
    &=
        \frac{
            \Pu
            \frac{\beta_{llk}}{2}
            \sum_{i=1}^{2\left(N-K+1\right)}
            Z_i^2
            }{
            \Pu
            \sum_{i \neq l}^{L}
                    \B{w}_i^H
                    \B{D}_{li}
                    \B{w}_i
            +
            1
            }.
\end{align}
By dividing the numerator and denominator of $\gamma_k$ in
\eqref{eq Asy 3d} by $2 \left(N-K+1\right)$, we obtain
\begin{align} \label{eq Asy 3d1}
    \gamma_k
    &=
        \frac{
            \Pu
            \frac{\beta_{llk}}{2}
            \sum_{i=1}^{2\left(N-K+1\right)}
            Z_i^2/\left(2 \left(N-K+1\right)\right)
            }{
            \left(\Pu
            \sum_{i \neq l}^{L}
                    \B{w}_i^H
                    \B{D}_{li}
                    \B{w}_i
            +
            1\right)/\left(2 \left(N-K+1\right)\right)
            }.
\end{align}
Since $N/K = \kappa$, $2 \left(N-K+1\right) = 2
\left(N-\frac{1}{\kappa} N+1\right) \to \infty$ as $N \to \infty$.
Thus, by using the law of large numbers and the trace lemma from
\cite[Lemma~13]{Hoy:12:Thesis}, i.e.,\footnote{Note that the trace
lemma holds if $\limsup_K \E \left\{\frac{1}{K} {\tt Tr}
\left(\B{D}_{li} \B{D}_{li}^H \right)^2 \right\}<\infty$ which is
equivalent to $\E \left\{\beta_{lik}^4 \right\}<\infty$
\cite[Remark~3]{Hoy:12:Thesis}. For example, if $\beta_{lik}$ is a
log-normal random variable with standard deviation of $\sigma$,
then $\E \left\{\beta_{lik}^4 \right\} = e^{8 \sigma^2}$
\cite{SY:82:BST}. Evidently, for the vast majority of practical
cases of interest, the standard deviation is finite, which makes
the fourth moment bounded.
        }
$$
    \frac{1}{K}
                    \B{w}_i^H
                    \B{D}_{li}
                    \B{w}_i
    -
    \frac{1}{K}
    {\tt Tr} \B{D}_{li}
    \mathop \rightarrow \limits^{\tt a.s.}
    0, ~ \text{as $K \to \infty$}
$$
we obtain
\begin{align} \label{eq Asy 3e1}
    \gamma_k
    -
    \frac{
        \beta_{llk}\left(\kappa-1 \right)
        }{
        \sum_{i=1,i \neq l}^{L} \frac{1}{K}{\tt Tr} \B{D}_{li}
        }
    \mathop \rightarrow \limits^{\tt a.s.} 0,  ~
        \text{as $N\to \infty$, and $N/K = \kappa$}.
\end{align}
Therefore a deterministic approximation,  $\bar{\gamma}_k$, of
$\gamma_k$ is given by
\begin{align} \label{eq Asy 3e}
    \bar{\gamma}_k
    =
    \frac{
        \beta_{llk}\left(\kappa-1 \right)
        }{
        \sum_{i=1,i \neq l}^{L} \frac{1}{K}{\tt Tr} \B{D}_{li}
        }.
\end{align}

It is interesting to note that the SIR expression \eqref{eq Asy
3e} is independent of the transmit power, and increases
monotonically with $\kappa$. Therefore, for an arbitrarily small
transmit power, the SIR \eqref{eq Asy 3e} can be approached
arbitrarily closely by using a sufficiently large number of
antennas and users. The reason is that since the number of users
$K$ is large, the system is interference-limited, so if every user
reduces its power by the same factor then the limiting SIR is
unchanged. Furthermore, from \eqref{eq Asy 3e}, when $\kappa \to
\infty$ (this is equivalent to the case $N \gg K$), the SIR
$\bar{\gamma}_k \to \infty$, as $N \to \infty$, which is
consistent with \eqref{eq Asy 3}.

\subsubsection{Fixed $N\Pu$, $N \rightarrow \infty$}
Let $\Pu = \Eu/N$, where $\Eu$ is fixed. From \eqref{eq Asy 3},
we have
\begin{align} \label{eq Asy 3b}
    \gamma_k
    &=
        \frac{
            \Eu
            \frac{\beta_{llk}}{2}
            \frac{
            \sum_{i=1}^{2\left(N-K+1\right)}
            Z_i^2
                }{
                2\left(N-K+1\right)
                }
            \frac{
                2\left(N-K+1\right)
                }{
                N
                }
            }{
            \frac{\Eu}{N}
            \sum_{i \neq l}^{L}
                \left\|
                    \B{Y}_i
                \right\|^2
            +
            1
            }.
\end{align}
Then, again using the law of large numbers and the trace lemma, we
obtain
%%
%\begin{align} \label{eq Asy 4}
%    \gamma_k
%    &\mathop \rightarrow \limits^{\tt a.s.}
%    \left\{%
%\begin{array}{l}
%  \beta_{llk} \Eu, \hspace{2.5cm}~ \text{as} ~ N \rightarrow \infty, ~ \text{and fixed $K$ }
%  \\
%  \frac{
%    \beta_{llk} \Eu \left(1-1/\kappa\right)
%    }{
%    \Eu/\kappa
%    \sum_{i=1, i\neq l}^{L}
%        \frac{1}{K}
%        {\tt Tr} \B{D}_{li}
%    +
%    1
%    }, \hspace{0.1cm}~ \text{as} ~ N \rightarrow \infty, ~ N/K=\kappa.\\
%\end{array}%
%\right.
%\end{align}
%
%
\begin{align} \label{eq Asy 4}
    &\gamma_k
    -
    \beta_{llk} \Eu
    \mathop \rightarrow \limits^{\tt a.s.} 0, \hspace{3.65cm}~ \text{as} ~ N \rightarrow \infty, ~ \text{and fixed $K$ }
    \\ \label{eq Asy 4b}
    &\gamma_k
    -
      \frac{
    \beta_{llk} \Eu \left(1-1/\kappa\right)
    }{
    \Eu/\kappa
    \sum_{i=1, i\neq l}^{L}
        \frac{1}{K}
        {\tt Tr} \B{D}_{li}
    +
    1
    }
    \mathop \rightarrow \limits^{\tt a.s.}
    0, \hspace{0.1cm}~ \text{as} ~ N \rightarrow \infty, ~ N/K=\kappa.
\end{align}
These results show that by using a very large antenna array at the
BS, we can cut the transmit power at each user proportionally to
$1/N$ while maintaining a desired quality-of-service. This result
was originally established in \cite{NLM:11:TCOM} for the case when
$N \gg K \gg 1$ whereas herein, we have generalized this result to
the regime where $N \gg 1$. Again, we can see that, when $\kappa$
tends to infinity, the two asymptotic results \eqref{eq Asy 4} and
\eqref{eq Asy 4b} coincide.

\begin{remark}\label{remark 2}
We can see from \eqref{eq Asy 4} that when $N$ grows without bound
and $K$ is fixed, the effects of interference and fast fading
disappear. The only remaining effect is noise. Let us define the
``massive MIMO effect" as the case where the system is ultimately limited
by noise.\footnote{
    The term ``massive MIMO effect'' was also used in
    \cite{HBD:11:ACCCC} but in a different meaning, namely referring to the case when the system performance is limited
    by pilot contamination, due to the use of non-orthogonal pilots in different cells for the uplink training phase.
    However, here we assume perfect CSI, and
    we are considering a particular operating condition where the
    transmit power is very small ($\Pu \sim 1/N$).}
From \eqref{eq Asy 4b}, when $N$ grows large while keeping a
finite $\kappa$, the system is still limited by interference from
the other cells. This interference depends mainly on $\kappa$ (the
degrees of freedom), and when $\kappa \to \infty$, we operate
under massive MIMO conditions. Therefore, an interesting question
is: How many degrees of freedom $\kappa$ are needed in order to
make the interference small compared to the noise (i.e., to reach
the massive MIMO condition)? Mathematically, we seek to find
$\kappa$ that satisfies
\begin{align} \label{eq: DoF 1}
        \log_2
        \left(
        1
        +
  \frac{
    \beta_{llk} \Eu \left(1-1/\kappa\right)
    }{
    \Eu/\kappa
    \sum_{i=1, i\neq l}^{L}
        \frac{1}{K}
        {\tt Tr} \B{D}_{li}
    +
    1
    }
        \right)
    \geq
    \eta
    R_{k,\infty}, ~ \text{for a desired $\eta \in \left(0,
    1\right)$}
\end{align}
where $R_{k,\infty} = \log_2\left(1 + \beta_{llk}\Eu\right)$ is
the ultimate rate which corresponds to the
    regime where $N \gg K \gg 1$. We more closely address this fundamental issue via simulations in Section~\ref{Sec:Results}.

\end{remark}

\begin{remark} \label{remark 3}
When $N \gg K \gg 1$ and $\Pu = \Eu/N$, using the property
$\psi(x)=\ln(x)+1/x+{\cal O}(1/x^2)$, and observing that the
second term of the exponential function approaches zero, we can
simplify (\ref{eq PropBound 1}) to get
%\begin{align}
$\left\langle R_k \right\rangle\geq \log_2\left(1+\beta_{llk}
\Eu\right),$
%\end{align}
which coincides with (\ref{eq Asy 4}). This implies that the
proposed lower bound becomes exact in the large-number-of-antennas
regime.
\end{remark}

%\vspace{-0.5cm}

\section{Numerical Results} \label{Sec:Results}

%\vspace{-0.4cm}

In this section, we provide some numerical results to verify our
analysis. Firstly, we consider a simple scenario where the
large-scale fading is fixed. This setting enables us to validate
the accuracy of our proposed analytical expressions as well as
study the fundamental effects of intercell interference, number of
BS antennas, transmit power of each user on the system
performance. We then consider a more practical scenario that
incorporates small-scale fading and large-scale fading including
path-loss, shadowing, and random user locations.

\vspace{-0.5cm}

\subsection{Scenario I}

\begin{figure}[t]
    \centering
    \centerline{\includegraphics[width=0.8\textwidth]{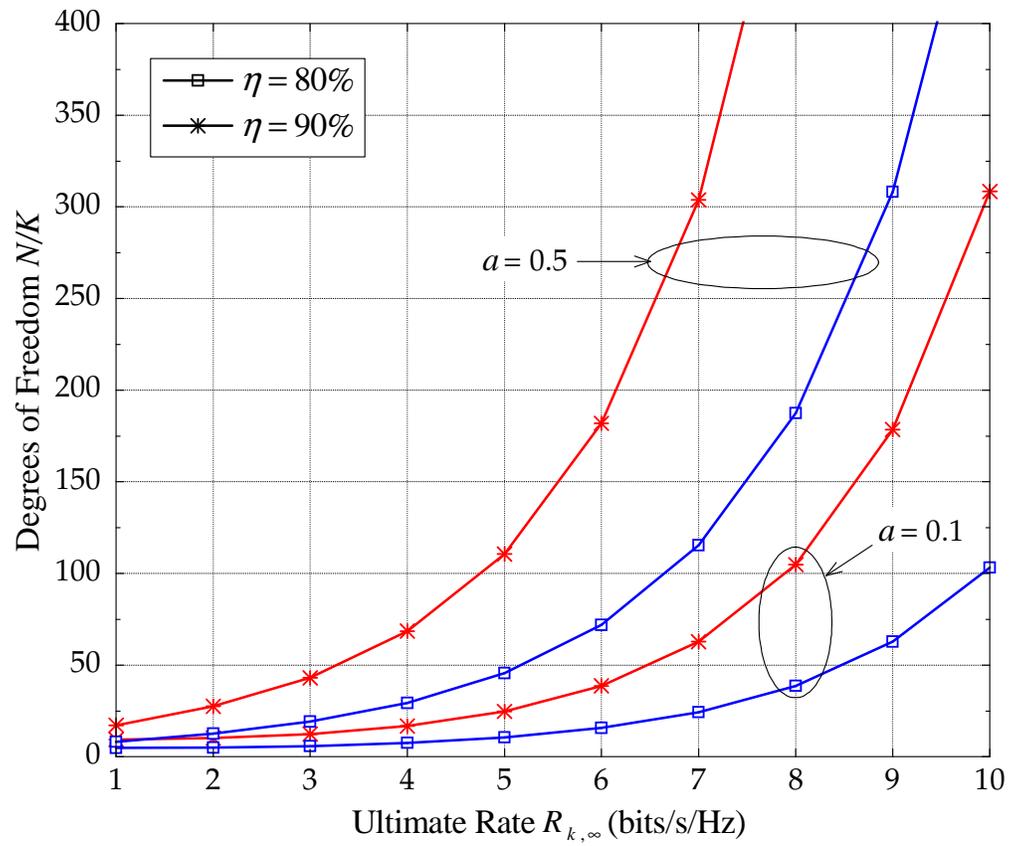}}
    \caption{Degrees of freedom $\kappa$ required to achieve $\eta R_{k, \infty}$ versus $R_{k, \infty}$ ($L=4$, $a=0.1$ and $0.5$).}
    \label{fig:5a}
\end{figure}

We consider a multicell MU-MIMO system with $4$ cells sharing the
same frequency band. In all examples, except Fig.~\ref{fig:5a}, we
choose the number of users per cell to be $K = 10$. We assume that
all direct gains are equal to $1$ and all cross gains are equal to
$a$, i.e., $\beta_{llk} = 1$, and $\beta_{ljk} = a, \forall j \neq
l$, $k = 1, 2, ..., K$ ($a$ can be regarded as an interference
factor). Furthermore, we define $\mathsf{SNR} \triangleq \Pu$.

\begin{figure}[t]
    \centerline{\includegraphics[width=0.8\textwidth]{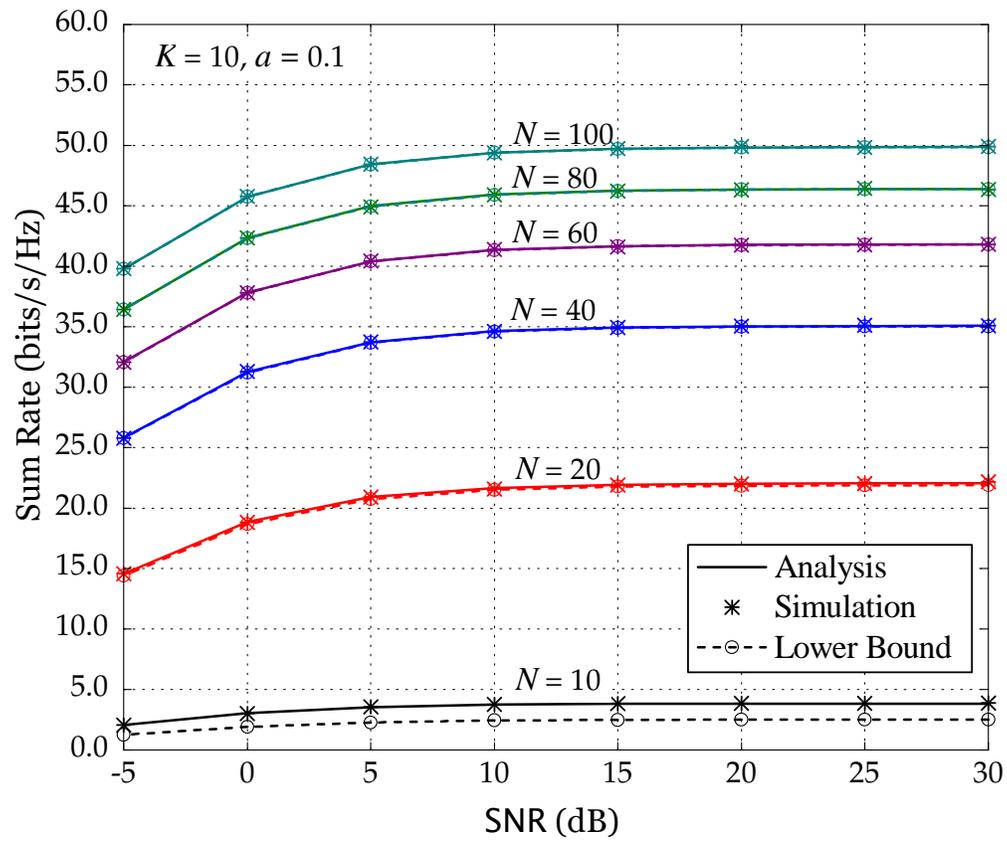}}
    \caption{Simulated uplink sum rate, analytical expression and lower bound versus the SNR ($L = 4$, $K=10$, and $a =0.1$).}
    \label{fig:1}
\end{figure}

Figure~\ref{fig:1} shows the uplink sum rate per cell versus
$\mathsf{SNR}$, at cross gain $a = 0.1$ and for different numbers
of BS antennas $N = 10$, $20$, $40$, $60$, $80$ and $100$. The
simulation curves are obtained by performing Monte-Carlo
simulations using \eqref{eq MU-MIMO 4}, while the analytical and
bound curves are computed via \eqref{eq Prop2 1} and \eqref{eq
PropBound 1}, respectively.  We can see that the simulated and
analytical results match exactly. As expected, when $N$ increases,
the sum rate increases too. However, at high SNRs, the sum rate
converges to a deterministic constant which verifies our analysis
\eqref{eq Asy 1}. Furthermore, a larger value of $N$ makes the
bound tighter. This is due to the fact that when $N$ grows large,
things that were random before become deterministic and hence,
Jensen's inequality used in \eqref{Rate 1a} will hold with
equality (see Remark~\ref{remark 3}). Therefore, the bound can
very efficiently approximate the rate when $N$ is large. It can be
also seen that, even for moderate number of antennas ($N
\gtrapprox 20$), the bound becomes almost exact across the entire
SNR range.

\begin{figure}[t]
    \centering
    \centerline{\includegraphics[width=0.8\textwidth]{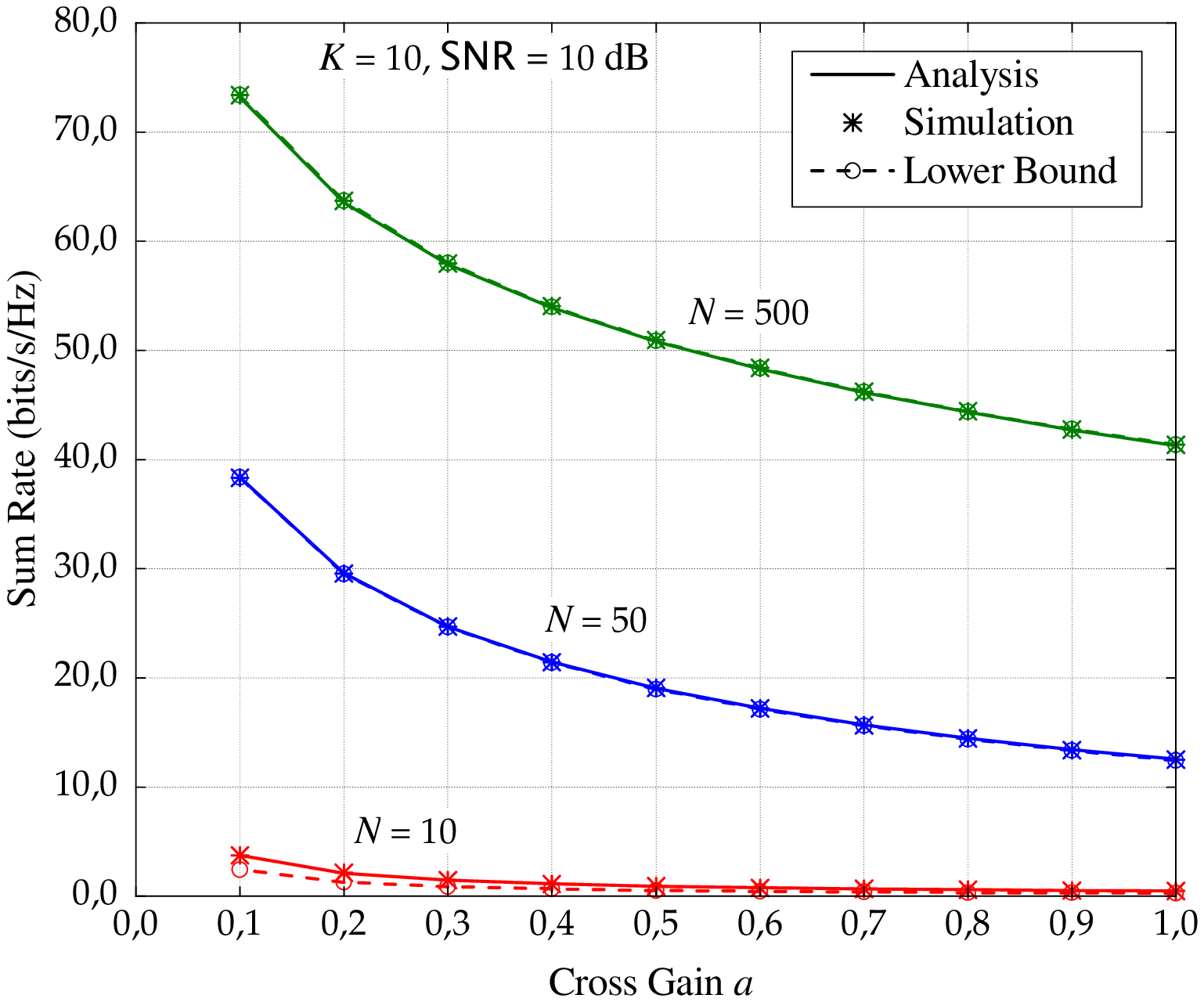}}
    \caption{Simulated uplink sum rate, analytical expression and lower bound versus the cross gain $a$ ($L = 4$, $K=10$, and $\mathsf{SNR} =10$ dB).}
    \label{fig:2}
\end{figure}

The effect of interference for different $N$ is shown in Fig.~\ref{fig:2}. Again, the simulated and
analytical results match exactly, and the bound is very tight.
Interestingly, its tightness does not depend on the interference
level but on the number of BS antennas. We can see that when the
cross gain increases (and hence, the interference increases), the
sum rate decreases significantly. On the other hand, the effect of
interference decreases when the number of BS antennas grows large.
For example, at $a=0.1$, the sum rates are $3.76$, $38.35$, and
$73.20$ for $N = 10$, $50$, and $500$, respectively, while at $a =
0.5$, the sum rates are respectively $0.93$, $19.10$, and $50.80$
for $N = 10$, $50$, and $500$. This means that when increasing the
cross gain from $0.1$ to $0.5$, the sum rates are reduced by
$75.27\%$, $50.20\%$, and $30.60\%$  for $N = 10$, $50$, and
$500$, respectively.

\begin{figure}[t]
    \centering
    \centerline{\includegraphics[width=0.8\textwidth]{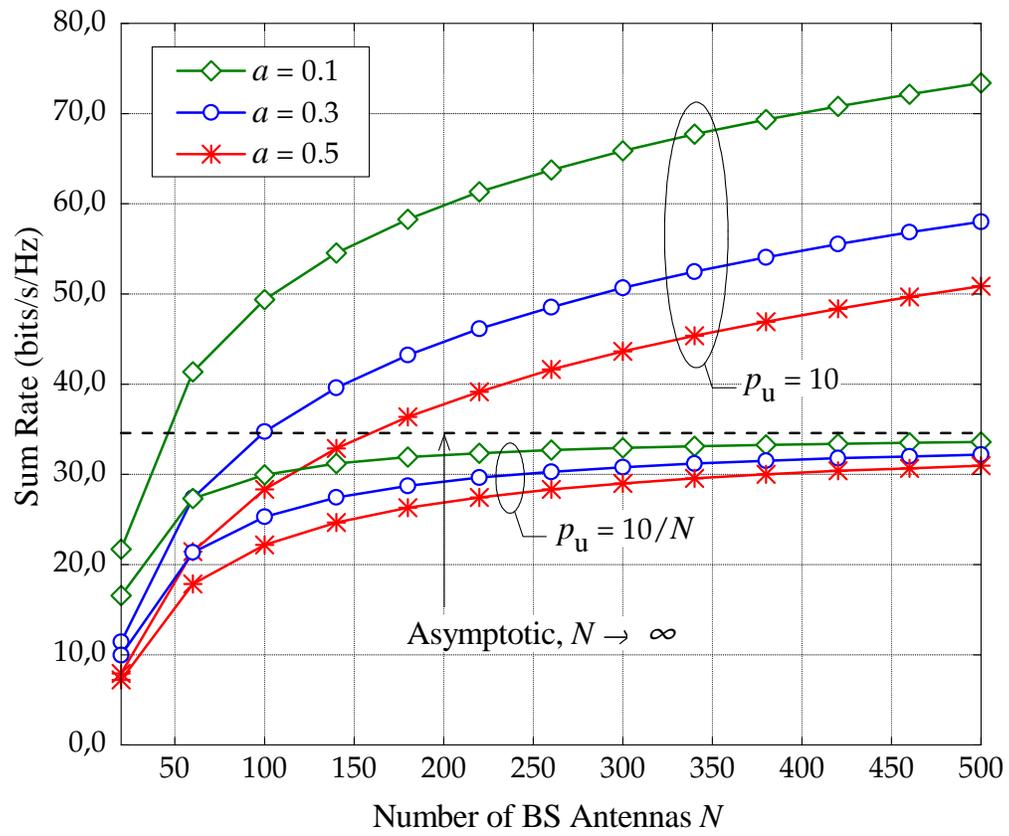}}
    \caption{Analytical uplink sum rate versus the number of BS antennas $N$ ($L=4$, $K=10$, $a=0.1$, $0.3$, and $0.5$).}
    \label{fig:3}
\end{figure}

The power efficiency of large   array systems is investigated in
Fig.~\ref{fig:3}. Figure~\ref{fig:3} shows the uplink sum rate per
cell versus $N$ at $a=0.1$, $0.3$, and $0.5$ for the cases of $\Pu
= 10$ and $\Pu = 10/N$. As expected, with $\Pu = 10/N$, the sum
rate converges to a constant value when $N$ increases regardless
of the effects of interference, and with $\Pu =10$, the sum rate
grows without bound (logarithmically fast with $N$) when $N$
increases (see \eqref{eq Asy 3} and \eqref{eq Asy 4}).

%We next consider the accuracy of the deterministic approximation
%\eqref{eq Asy 3e}. Figure.~\ref{fig:4} shows the sum rate as a
%function of the number of BS antennas $N$ for $N/K = 2$,
%$\mathsf{SNR} = 10$ dB, and for   interference factors $a=0.1$,
%$a=0.3$, and $a=0.5$. As expected, the deterministic approximation
%is very close to the exact analysis even when $N$ is not so large
%($N \thickapprox 10$). Keeping a finite ratio $N/K$ makes the sum
%rate increase linearly with $N$. Furthermore, the performance
%degrades significantly with a stronger interference.

Figure~\ref{fig:5a} shows the required number of degrees of
freedom $\kappa$ to achieve $80 \%$ ($\eta = 0.8$) and  $90 \%$
($\eta = 0.9$) of a given ultimate rate $R_{k, \infty}$, for
$a=0.1$, and $a=0.5$. We use \eqref{eq: DoF 1} to determine
$\kappa$. We can see that $\kappa$ increases with $R_{k, \infty}$.
Therefore, for multicell systems, the BS can serve more users with
low data rates. This is due to the fact that when $R_{k, \infty}$
increases, the transmit power increases and hence, the
interference also increases. Then, we need more degrees of freedom
to mitigate interference. For the same reason, we can observe that
when the interference factor $a$ increases, the required $\kappa$
increases as well.

\begin{figure}[t]
    \centering
    \centerline{\includegraphics[width=0.8\textwidth]{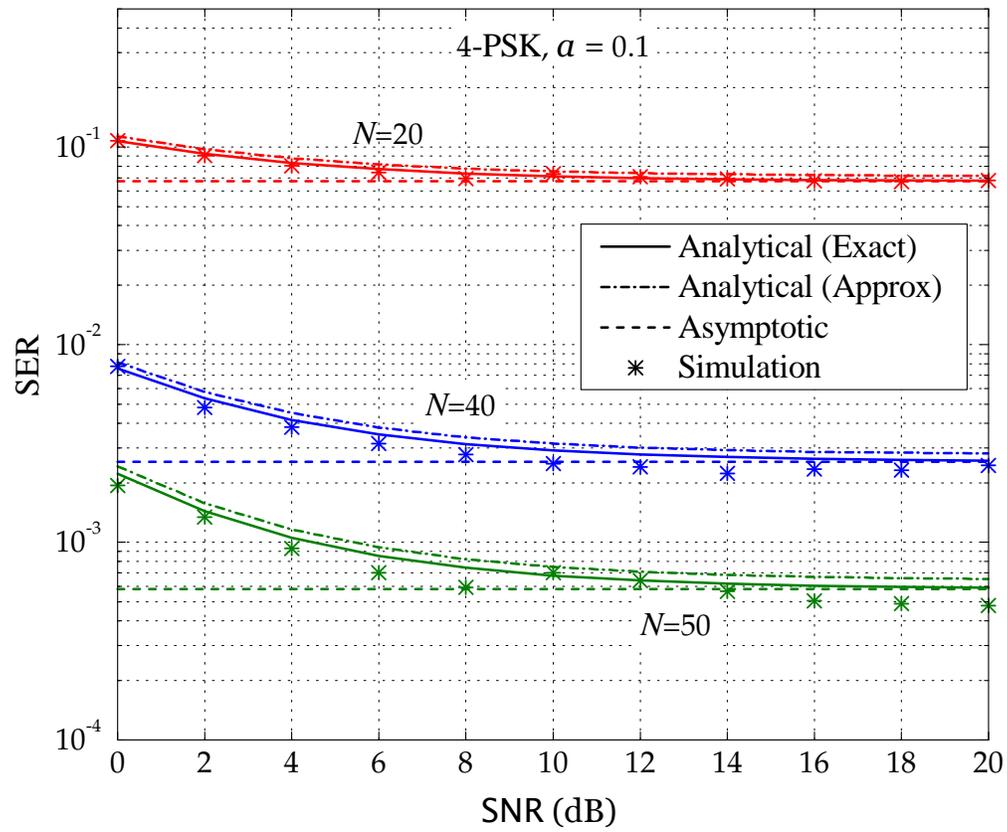}}
    \caption{Simulated average SER and analytical expression versus the $\mathsf{SNR}$ for $4$-PSK ($L=4$, $K=10$ and $a=0.1$).}
    \label{fig:5}
\end{figure}

\begin{figure}[t]
    \centering
    \centerline{\includegraphics[width=0.8\textwidth]{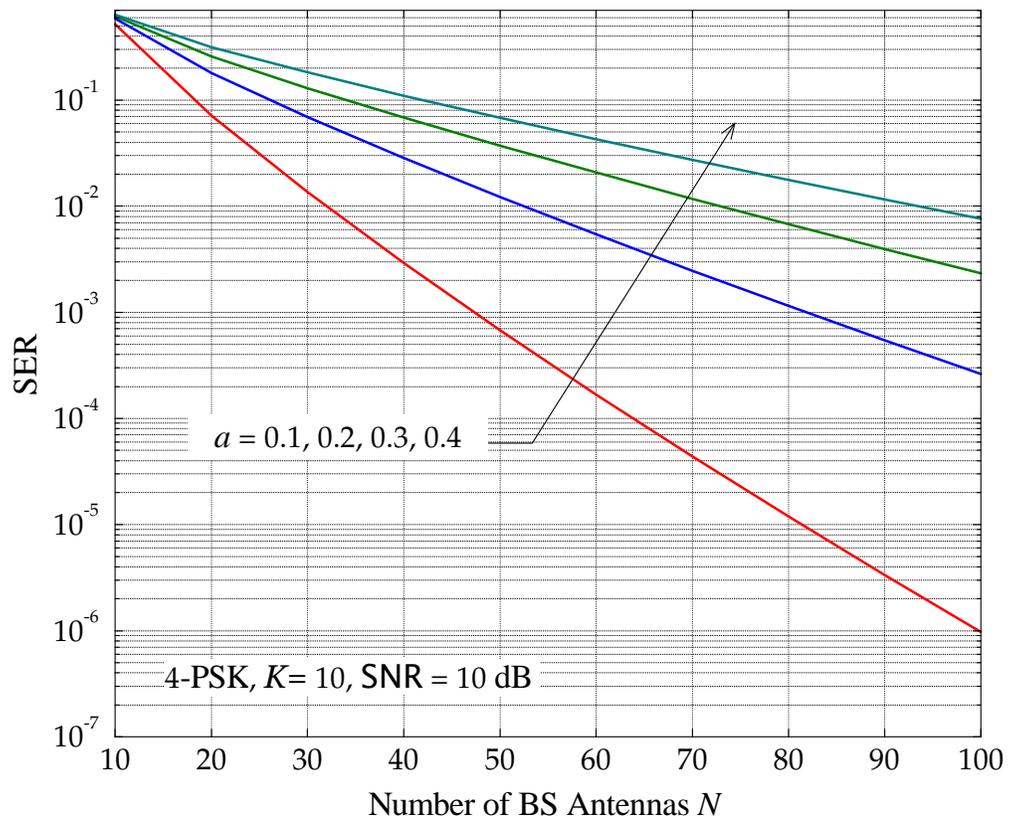}}
    \caption{Analytical average SER versus the number of BS antennas $N$ ($L=4$, $K=10$, $a=0.1$, $0.2$, $0.3$, and $0.4$).}
    \label{fig:6}
\end{figure}

In Fig.~\ref{fig:5}, the analytical SER curves are compared with
the outputs of a Monte-Carlo simulation for different $N$.  Here,
we choose $4$-PSK and $a = 0.1$. The ``Analytical (Exact)" curves
are computed using Theorem~\ref{prop: SER}, and the ``Analytical
(Approx)" curves are generated using \eqref{eq SER 2c}. In
addition, the high-SNR curves, generated via \eqref{eq SER 1a},
are also overlaid. It can be easily observed that the analytical
results coincide with the simulation results. Furthermore, we can
see that the ``Analytical (Approx)" curves are accurate in all
cases. As in the analysis of the sum rate, when the SNR is
moderately large, the SER decreases very slowly and approaches an
error floor (the asymptotic SER) due to interference, when SNR
grows large. Yet, we can improve the system performance by
increasing the number of BS antennas. The effects of using large
antenna arrays on the SER can be further verified in
Fig.~\ref{fig:6}, where the SER is plotted as a function of $N$
for different cross gains and $4$-PSK, at $\mathsf{SNR} = 10$ dB.
We can see that the system performance improves systematically
when we increase $N$.

%Figure~\ref{fig:7} depicts the outage probability of each user
%against the number of BS antennas $N$ for a fixed threshold value
%$\gamma_{\tt th} = 0$ dB (corresponding to an achievable rate of
%$1$ bits/s/ Hz per user) and different $a$, at $\mathsf{SNR} = 0$
%dB. Once again, there is an exact agreement between the analytical
%results (using \eqref{eq Pro Out 1}) and simulated results. We can
%see again, as expected, that the number of BS antennas has a very
%strong impact on performance.

%Finally, Fig.~\ref{fig:8} investigates the outage probability
%versus the data rate of each user for $N = 20$, $40$, $60$, $80$,
%and $100$, at $a=0.1$, and $\mathsf{SNR} = 0$ dB. For a given
%indicated rate, the outage probability significantly decreases
%when $N$ increases. The circles correspond to $95$\%-likely rates,
%i.e., the rate is greater than or equal to this indicated rate
%with probability $0.95$. We can see that $95$\%-likely rates
%increase with $N$, i.e., they are equal to $1.30, 2.64, 3.34,
%3.83$, and $4.19$ when $N = 20, 40, 60, 80$, and $100$.
%Furthermore, when $N$ is large, the random system becomes
%deterministic and hence, the probability that the uplink rate is
%around its mean is inherently high.
\vspace{-0.5cm}
\subsection{Scenario II}

We consider a hexagonal cellular network where each cell has a
radius (from center to vertex) of $1000$ meters. In each cell,
$K=10$ users are located uniformly at random and we assume that no
user is closer to the BS than $r_{\mathrm{h}}= 100$ meters. The
large-scale fading is modeled via $\beta_{lik}=
z_{lik}/\left(r_{lik}/r_{\mathrm{h}} \right)^\nu$, where $z_{lik}$
represents a log-normal RV with standard deviation of $8$ dB,
$r_{lik}$ is the distance between the $k$th user in the $i$th cell
to the $l$th BS, and $\nu$ is the path loss exponent. We choose
$\nu = 3.8$ for our simulations. Furthermore, we assume that the
transmitted data is modulated using OFDM. Let $T_{\mathrm{s}}$ and
$T_{\mathrm{u}}$ be the OFDM symbol duration and useful symbol
duration, respectively. Then, we define the net uplink rate of the
$k$th user in the $l$th cell as follows \cite{Mar:10:WCOM}:
\begin{align} \label{eq: net rate 1}
    R_k^{\mathrm{net}}
    =
    \frac{B}{r}
    \frac{T_{\mathrm{u}}}{T_{\mathrm{s}}}
    \log_2
        \left(
        1
        +
        \frac{
            \Pu X_k
            }{
            \Pu Z_l
            +1/r
            }
        \right)
\end{align}
where $B$ is the total bandwidth, and $r$ is the frequency-reuse
factor. Note that \eqref{eq: net rate 1} is obtained by using the
result in Proposition~\ref{Prop 1}. For our simulations, we choose
parameters that resemble those of the LTE standard
\cite{Mar:10:WCOM}: $T_{\mathrm{s}}=71.4 \mu$sec, and
$T_{\mathrm{s}}=66.7 \mu$sec. We further assume that the total
bandwidth of the system is $20$ MHz.
 We neglect the effects of all
users in all cells which are outside a circular region with a
radius (from the $l$th BS) of $8000$ meters. This is reasonable
since the interference from all users which are outside this
region is negligible due to the path loss.

\begin{figure}[t]
    \centering
    \centerline{\includegraphics[width=0.8\textwidth]{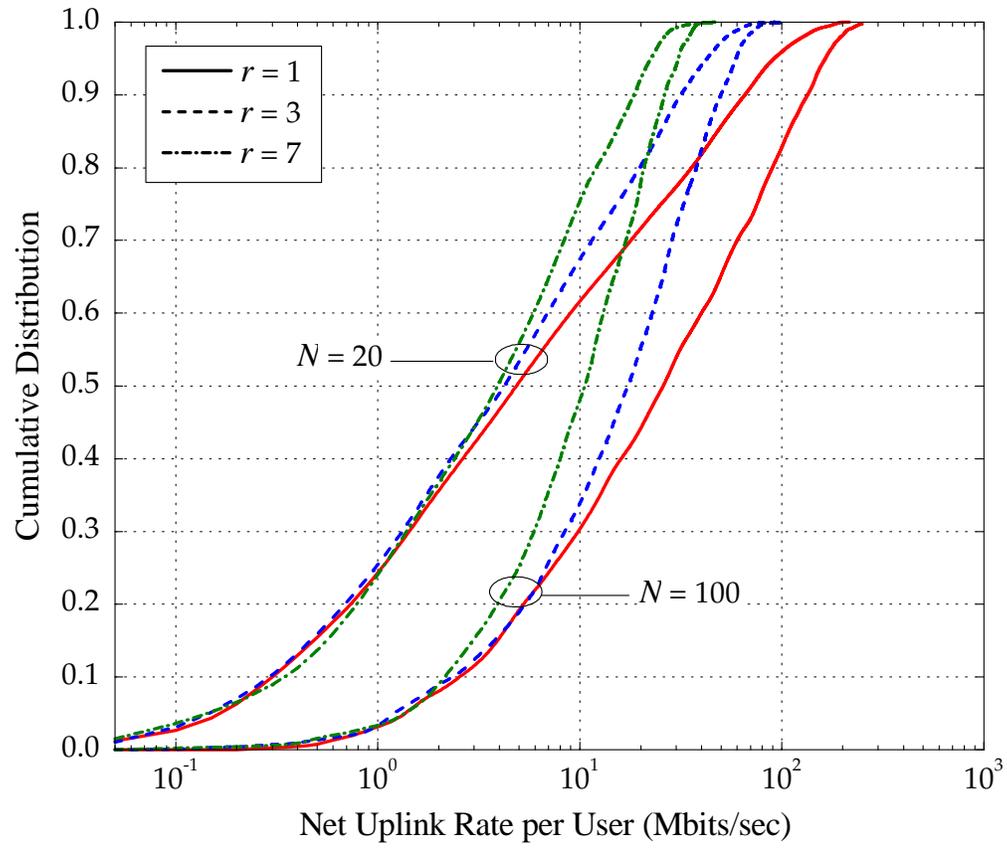}}
    \caption{Cumulative distribution of the net uplink rate per user for frequency-reuse factors $1$, $3$, and $7$
    ($N=20$, $100$, $\mathsf{SNR} = 10$ dB, $\sigma_{\mathrm{shadow}}=8$ dB, and $\nu = 3.8$).}
    \label{fig:9}
\end{figure}

Figure~\ref{fig:9} shows the cumulative distribution of the net
uplink rate per user for different frequency-reuse factors $r=1$,
$3$, and $7$, and different number of BS antennas $N=20, 100$. We
can see that the number of BS antennas has a very strong impact on
the performance. The probability that the net uplink rate is
smaller than a given indicated rate decreases significantly when
$N$ increases. We consider the $95$\%-likely rates, i.e., the rate
is greater than or equal to this indicated rate with probability
$0.95$. We can see that $95$\%-likely rates increase with $N$; for
example, with frequency-reuse factor of $1$, increasing the number
of BS antennas from $20$ to $100$ yields a $8$-fold improvement in
the $95$\%-likely rate (from $0.170$ Mbits/sec to $1.375$
Mbits/sec). Furthermore, when $N$ is large, the random channel
becomes deterministic and hence, the probability that the uplink
rate is around its mean becomes inherently higher.

When comparing the effects of using frequency-reuse factors, we
can see that, at high rate (and hence at high SNR), smaller reuse
factors are preferable, and vice versa at low rate. Furthermore,
we can observe that the gap between the performance of different
reuse factors becomes larger when $N$ increases. This is due to
the fact that, when $N$ is large, the intercell interference can
be notably reduced; as a consequence, the bandwidth used has a
larger impact on the system performance. Table~\ref{table:1}
summarizes the $95$\%-likely net uplink rates as well as their
mean values.

\section{Conclusion} \label{Sec:Conclusion}
In this paper, we analyzed in detail the uplink performance of
data transmission from $K$ single-antenna users in one cell to its
$N$-antenna BS in the presence of interference from other cells.
The BS uses ZF to detect the transmitted signals. We derived exact
closed-form expressions for the most important figures of merit, namely the uplink rate, SER, and outage
probability, assuming that the channel between the users and the
BS is affected by Rayleigh fading, shadowing, and path loss.

Theoretically, when $N$ increases we obtain array and diversity
gains, which affect both the
 desired and interference signals.
Hence, from this perspective the performance is not dramatically
affected. However, since when the number of BS antennas is large,
the channel vectors between the users and the BS are pairwisely
asymptotically orthogonal, the interference can be cancelled out
with a simple linear ZF receiver. As a consequence, by using a
large antenna array, the performance of the multicell system
improves significantly. Furthermore, we investigated the
achievable power efficiency when using large antenna arrays at the
BSs. Large antenna arrays enable us to reduce the transmitted
power of each user proportionally to $1/N$ with no performance
degradation, provided that the BS has perfect CSI. We further
elaborated on the massive MIMO effect and the impact of
frequency-reuse factors.

\appendix

\subsection{Proof of Theorem~\ref{Prop 2}} \label{sec:proof:prop rate1}

We first present the following lemma which is used to derive the
closed-form expression of the achievable ergodic rate in
Section~\ref{Sec:Rate_CF}.

\begin{lemma}\label{lemma1}
Let $m, n \in \mathbb{N}$, $a, b \in \mathbb{R}$, and $\alpha \in
\mathbb{R}^{+}$. Then
\begin{align}\label{eq lemma1 1}
    \int_{0}^{\infty}
        x^m
        \left(ax +b\right)^n
        e^{-\alpha x}
        \mathrm{Ei} \left( -\left(ax+b\right)\right)
        dx
    =
        \mathcal{I}_{m,n}\left(a,b,\alpha\right)
\end{align}
where $\mathcal{I}_{m,n}\left(a,b,\alpha\right)$ is given by
\eqref{I Func}.
\begin{proof}
By applying the change of variables $z = ax+b$ in \eqref{eq lemma1
1}, we have
\begin{align}\label{eq prooflemma1 1}
    \int_{0}^{\infty}
        x^m
        \left(ax +b\right)^n
        e^{-\alpha x}
        \mathrm{Ei} \left( -\left(ax+b\right)\right)
        dx
%    \nonumber
%    \\
%    &\hspace{1.7cm}
    &=
    \frac{e^{\frac{\alpha b}{a}}}{a^{m+1}}
    \int_{b}^{\infty}
        \left(z-b \right)^m
        z^n
        e^{- \frac{\alpha z}{a}}
        \mathrm{Ei} \left( -z\right)
        dz
    \nonumber
    \\
    &=
    \frac{e^{\frac{\alpha b}{a}}}{a^{m+1}}
    \sum_{i=0}^{m}
    \binom{m}{i}
        \left(-b\right)^{m-i}
    \mathcal{J}_{n+i}\left(b,\frac{\alpha}{a}\right)
\end{align}
where
\begin{align} \label{eq:proof coro1 9}
    \mathcal{J}_p\left(\alpha,\mu\right)
    \triangleq
        \int_\alpha^\infty
            z^p
            e^{-\mu z}
            \mathrm{Ei}\left(-z\right)
            dz,
            \quad
            \mu \geq 0.
\end{align}
Using  partial integration and the fact that $d
\mathrm{Ei}\left(z\right)/dz =e^z/z$, the integral \eqref{eq:proof
coro1 9} for $p \in \mathbb{N}$ can be evaluated as
\begin{align}\label{eq:proof coro1 13}
    \mathcal{J}_p\left(\alpha,\mu\right)
    &=
        -\left.
            \frac{e^{-\mu z} z^p}{\mu}
            \mathrm{Ei}\left(-z\right)
        \right|_\alpha^\infty
%    \nonumber \\
%    &~ ~ ~
        +
        \frac{1}{\mu}
        \int_\alpha^\infty
            e^{-\mu z}
            \left[
                p z^{p-1}
                \mathrm{Ei}\left(-z\right)
                +
                z^{p-1}
                e^{-z}
            \right]
            dz
    \nonumber \\
    &=
        \mathcal{K}_p\left(\alpha,\mu\right)
        +
        \frac{p}{\mu}
        \mathcal{J}_{p-1}\left(\alpha,\mu\right)
    %\nonumber \\
    =
        \sum_{q=0}^{p-1}
            \left(
                \tfrac{p}{\mu}
            \right)^q
            \mathcal{K}_{p-q}\left(\alpha,\mu\right)
        +
        \left(
            \tfrac{p}{\mu}
        \right)^p
        \mathcal{J}_{0}\left(\alpha,\mu\right)
\end{align}
where the second equality follows from
\cite[Eq.~(2.321.2)]{GR:07:Book}, where
\begin{align} \label{eq:proof coro1 12}
    \mathcal{K}_p\!\left(\alpha,\mu\right)
    &\triangleq \!
        \frac{e^{-\alpha \mu}\alpha^p}{\mu}
        \mathrm{Ei}\left(\!-\alpha \!\right)
%    \nonumber \\
%    &\hspace{0.5cm}
        +
        \frac{e^{-\alpha\left(\mu+1\right)}}{\mu}
        \!\sum_{q=0}^{p-1}
            \frac{
                q!
                \binom{p-1}{q}
                \alpha^{p-q-1}
            }{
                \left(\mu+1\right)^{q+1}
            }
\end{align}
while the last equality is obtained using recursion. Finally, by
using \cite[Eqs.~(5.231.2) and (6.224.1)]{GR:07:Book},
$\mathcal{J}_{0}\left(\alpha,\mu\right)$ in \eqref{eq:proof coro1
13} can be evaluated as
\begin{align} \label{eq:proof coro1 10}
    \mathcal{J}_0\left(\alpha,\mu\right)
    =
        -\frac{1}{\mu}
        \mathrm{Ei}\left(-\left(\mu+1\right)\alpha\right)
        +\frac{e^{-\alpha \mu}}{\mu}
        \mathrm{Ei}\left(-\alpha\right).
\end{align}
From \eqref{eq prooflemma1 1}, \eqref{eq:proof coro1
13}--\eqref{eq:proof coro1 10}, we arrive at the desired result
\eqref{eq lemma1 1}.
\end{proof}
\end{lemma}

Using Lemma~\ref{lemma1}, and \cite[Eq.~(39)]{KA:06:WCOM}, we
can easily obtain \eqref{eq Prop2 1}.

\subsection{Proof of Proposition~\ref{Prop BoundRate 1}}\label{sec:proof:propBound Rate}
From \eqref{Rate 1}, the uplink ergodic rate from the $k$th user
in the $l$th cell to its BS can be expressed as:
\begin{align} \label{Rate 1b}
    \left\langle R_k \right\rangle
&=\EXs{X_k, Z_l}{\log_2\left(1+\Pu\exp\left(\ln\left(\frac{
X_k}{\Pu Z_l + 1}\right)\right)\right)}
\\
&\geq\log_2\left(1+\Pu\exp\left(\EXs{X_k, Z_l}{\ln\left(\frac{ X_k}{\Pu Z_l + 1}\right)}\right)\right)   \label{Rate 1a}\\
&=\log_2\left(1+\Pu\exp\left(\EXs{X_k}{\ln\left(X_k\right)}-\EXs{Z_l}{\ln\left(\Pu
Z_l + 1\right)}\right)\right)  \label{eq:lowerbound}
\end{align}
where we have exploited the fact that $\log_2(1+\alpha \exp(x))$
is convex in $x$ for $\alpha>0$ along with Jensen's inequality. We
can now evaluate the expectations in (\ref{eq:lowerbound}) and we
begin with $\EXs{X_k}{\ln\left(X_k\right)}$, which can be
expressed as
\begin{align}
\EXs{X_k}{\ln\left(X_k\right)} =\frac{\beta_{llk}^{-N+K-1}}{\left(N-K\right)!
}\int_0^\infty
\ln(x)e^{-x/\beta_{llk}}x^{N-K}dx
%\nonumber
%\\
=\psi(N-K+1)+\ln(\beta_{llk}) \label{eq:firstexp}
\end{align}
where we have used \cite[Eq.~(4.352.1)]{GR:07:Book} to evaluate
the corresponding integral. The second expectation in
(\ref{eq:lowerbound}) requires a different line of reasoning. In
particular, we have that
\begin{align}
\EXs{Z_l}{\ln\left(\Pu Z_l + 1\right)}&= \sum_{m=1}^{\varrho\left(
\mathbf{\mathcal{A}}_l\right)} \sum_{n=1}^{\tau_m \left(
\mathbf{\mathcal{A}}_l\right)} \mathcal{X}_{m,n} \left(
\mathbf{\mathcal{A}}_l\right)
\frac{\mu_{l,m}^{-n}}{\left(n-1\right)!}
%\notag\\
%&\times
\underbrace{\int_0^\infty\ln\left(\Pu z +
1\right)z^{n-1}e^{\frac{-z}{\mu_{l,m}}}dz}_{\triangleq \cal I}.
\label{eq:secondexp}
\end{align}
The integral $\cal I$ admits the following manipulations
\begin{align}
{\cal I}&=\int_0^\infty G^{1,2}_{2,2}\left[ \Pu z \left\vert
\begin{array}{c}
1,1 \\
1,0 \\
\end{array}%
\right. \right]z^{n-1}e^{\frac{-z}{\mu_{l,m}}}dz
=\mu_{l,m}^nG^{1,3}_{3,2}\left[ \Pu\mu_{l,m}\left\vert
\begin{array}{c}
1-n,1,1 \\
1,0 \\
\end{array}%
\right. \right]\label{eq:Meijer1}
\end{align}
where $G \substack{ m, n \\
p , q } \left[ x, \left| \substack{ \alpha_1,\ldots,\alpha_p \\
\null \\ \beta_1,\ldots,\beta_q } \right. \right]$ denotes the
Meijer's-$G$ function~\cite[Eq.~(9.301)]{GR:07:Book}, and we have
expressed the integrand $\ln(1 + \alpha z)$ in terms of
Meijer's-$G$ function according to
\cite[Eq.~(8.4.6.5)]{PBM:90:Book:v3}. The final expression stems
from \cite[Eq.~(7.813.1)]{GR:07:Book}. In addition, we can
simplify (\ref{eq:Meijer1}) as follows
\begin{align}
{\cal I}= \Pu\mu_{l,m}^{n+1}G^{1,3}_{3,2}\left[
\Pu\mu_{l,m}\left\vert
\begin{array}{c}
-n,0,0 \\
0,-1 \\
\end{array}%
\right. \right]
%\nonumber\\
=\Pu\mu_{l,m}^{n+1}\Gamma(n+1){}_3F_1\left(n+1,1,1;2;-
\Pu\mu_{l,m}\right)\label{eq:Meijer3}
\end{align}
where we have used \cite[Eq.~(9.31.5)]{GR:07:Book} to obtain the
first equality and \cite[Eq.~(8.4.51.1)]{PBM:90:Book:v3} to obtain
the second equality. Combining (\ref{eq:secondexp}) with
(\ref{eq:Meijer3}) and after some basic simplifications, we get
\begin{align}
&\EXs{Z_l}{\ln\left(\Pu Z_l + 1\right)}=
\Pu\sum_{m=1}^{\varrho\left( \mathbf{\mathcal{A}}_l\right)}
\sum_{n=1}^{\tau_m \left( \mathbf{\mathcal{A}}_l\right)} \mu_{l,m}
n\mathcal{X}_{m,n} \left( \mathbf{\mathcal{A}}_l\right)
%\notag\\
%&\times
{}_3F_1\left(n+1,1,1;2;- \Pu\mu_{l,m}\right).
\label{eq:finalexp}
\end{align}
Substituting \eqref{eq:firstexp} and \eqref{eq:finalexp} into
\eqref{eq:lowerbound}, we conclude the proof.

\subsection{Proof of Theorem~\ref{prop: SER}}\label{sec:proof:prop SER}
The MGF of $\gamma_k$ is given by
\begin{align}\label{eq:proof:SER 1}
    \mathcal{M}_{\gamma_k}\left(s\right)
    &=
        \E_{\gamma_k} \left\{e^{-s \gamma_k} \right\}
%        \nonumber
%    \\
    =
        \int_{0}^{\infty}
        \E_{X_k} \left\{e^{-s \gamma_k} \right\}
        \PDF{Z_l}{z}
        d z.
\end{align}
Using the PDF of $X_k$ given by \eqref{eq PDF1 1}, we have that
\begin{align}\label{eq:proof:SER 2}
    \E_{X_k} \left\{e^{-s \gamma_k} \right\}
    &=
        \frac{1}{\left(N-K\right)! \beta_{llk}^{N-K+1}}
        \int_{0}^{\infty}
        x^{N-K}
        \exp \left(-x\left(\frac{1}{\beta_{llk}}+\frac{s}{z+1/\Pu}\right) \right)
        d x
    \nonumber
    \\
    &=
    \left(
        \frac{z + 1/\Pu}{z+1/\Pu+\beta_{llk} s}
    \right)^{N-K+1}
\end{align}
where the last equality is obtained by using
\cite[Eq.~(3.326.2)]{GR:07:Book}. Substituting \eqref{eq:proof:SER
2} into \eqref{eq:proof:SER 1} and using \eqref{eq PDF1 2}, we get
\begin{align}\label{eq:proof:SER 3}
    \mathcal{M}_{\gamma_k}\left(s\right)
    &=
        \int_{0}^{\infty}
        \sum_{m=1}^{\varrho\left( \mathbf{\mathcal{A}}_l\right)}
        \sum_{n=1}^{\tau_m \left( \mathbf{\mathcal{A}}_l\right)} \!
            \mathcal{X}_{m,n} \left( \mathbf{\mathcal{A}}_l\right)
            \frac{
                \mu_{l,m}^{-n}
            }{
                \left(
                    n-1
                \right)
                !
            }
            z^{n-1}
            e^{\frac{-z}{\mu_{l,m}}}
    \left(
        \frac{z + 1/\Pu}{z+ 1/\Pu +\beta_{llk} s}
    \right)^{N-K+1}
        d z
        \nonumber
        \\
    &=
        \sum_{m=1}^{\varrho\left( \mathbf{\mathcal{A}}_l\right)}
        \sum_{n=1}^{\tau_m \left( \mathbf{\mathcal{A}}_l\right)}
        \sum_{p=0}^{N-K+1}
        \binom{N-K+1}{p}
            \mathcal{X}_{m,n} \left( \mathbf{\mathcal{A}}_l\right)
            \frac{\left(-1\right)^{p}
                \mu_{l,m}^{-n}
            }{
                \left(
                    n-1
                \right)
                !
            }
        \left(\frac{\beta_{llk} s}{\beta_{llk}s +1/\Pu} \right)^{p}
    \nonumber
    \\
    &\times
        \int_{0}^{\infty}
            z^{n-1}
            e^{\frac{-z}{\mu_{l,m}}}
        \left(\frac{z}{1/\Pu + \beta_{llk}s} + 1\right)^{-p}
        d z
\end{align}
where the last equality is obtained by using the binomial
expansion formula. To evaluate the integral in \eqref{eq:proof:SER
3}, we first express $\left(\frac{z}{1/\Pu+\beta_{llk}s} +
1\right)^{-p}$ in terms of a Meijer's-$G$ function with the help
of \cite[Eq.~(8.4.2.5)]{GR:07:Book}, and then using the identity
\cite[Eq.~(2.24.3.1)]{GR:07:Book} to obtain
\begin{align}\label{eq:proof:SER 3b}
    \mathcal{M}_{\gamma_k}\left(s\right)
    &=
        \sum_{m=1}^{\varrho\left( \mathbf{\mathcal{A}}_l\right)}
        \sum_{n=1}^{\tau_m \left( \mathbf{\mathcal{A}}_l\right)}
        \sum_{p=0}^{M-K+1}
        \binom{N-K+1}{p}
            \mathcal{X}_{m,n} \left( \mathbf{\mathcal{A}}_l\right)
            \frac{\left(-1\right)^{p}
            }{
                \left(
                    n-1
                \right)
                !
            }
        \left(\frac{\beta_{llk} s}{\beta_{llk}s +1/\Pu} \right)^{p}
    \nonumber
    \\
    &\times
        \frac{1}{\Gamma(p)} G \substack{ 1, 2 \\
                2 , 1 }
                \left[ \frac{\mu_{l,m}}{\beta_{llk} s +1/\Pu} \left| \substack{ 1-n, 1-p \\
                    \null \\ 0 } \right.
                \right].
\end{align}
Finally, using \cite[Eq.~(8.4.51.1)]{PBM:90:Book:v3}, we arrive at
the desired result \eqref{eq SER 2}.

\subsection{Proof of Theorem~\ref{pro: out}} \label{sec:proof:Prop out}
From Proposition~\ref{Prop 1} and \eqref{eq Out 1}, we have
\begin{align}
 P_{{\tt out}}
 &\triangleq {\tt Pr}\left(\frac{X_k}{Z_l + 1/\Pu}\leq\gamma_{{\tt
 th}}\right).
\end{align}
We can now express the above probability in integral form as
follows:
\begin{align}\label{eq:CDF1}
P_{{\tt out}}=\int_0^\infty {\tt Pr}\left(X_k< \gamma_{{\tt
th}}(Z_\ell+1/\Pu)\left.\right|Z_\ell\right)p_{Z_\ell}(z)dz.
\end{align}
The cumulative density function (CDF) of $X_k$ can be shown to be equal to
\begin{align}
F_{X_k}(x)&={\tt Pr}\left(X_k\leq x\right)
=1-\exp\left(-\frac{x}{\beta_{llk}}\right)\sum_{p=0}^{M-K}\frac{1}{p!}\left(\frac{x}{\beta_{llk}}\right)^p
\label{eq:CDF2}
\end{align}
where we have used the integral identity
\cite[Eq.~(3.351.1)]{GR:07:Book} to evaluate the CDF. Combining
(\ref{eq:CDF1}) with (\ref{eq:CDF2}), we can rewrite $P_{{\tt
out}}$ as follows:
\begin{align}
    P_{{\tt out}}
    &=
        \int_0^\infty \left(1-\exp\left(-\frac{\gamma_{{\tt th}}(z+1/\Pu)}{\beta_{llk}}\right)\sum_{p=0}^{N-K}\frac{1}{p!}\left(\frac{\gamma_{{\tt th}}(z+1/\Pu)}{\beta_{llk}}\right)^p\right)p_{Z_\ell}(z)dz
       \nonumber \\
    &=
        1-\exp\left(-\frac{\gamma_{{\tt th}}}{\Pu\beta_{llk}}\right)\sum_{p=0}^{N-K}\frac{1}{p!}\int_0^\infty\exp\left(-\frac{\gamma_{{\tt th}}z}{\beta_{llk}}\right)\left(\frac{\gamma_{{\tt th}}(z+1/\Pu)}{\beta_{llk}}\right)^p
        p_{Z_\ell}(z)dz
       \nonumber \\
    &=
        1-\exp\left(-\frac{\gamma_{{\tt th}}}{\Pu\beta_{llk}}\right)\sum_{p=0}^{N-K}\frac{\left(\frac{\gamma_{{\tt th}}}{\beta_{llk}}\right)^p}{p!}\int_0^\infty\exp\left(-\frac{\gamma_{{\tt th}}z}{\beta_{llk}}\right)\left(1/\Pu+z\right)^p p_{Z_\ell}(z)dz
       \nonumber \\
    &=
        1-\exp\left(-\frac{\gamma_{{\tt
        th}}}{\Pu\beta_{llk}}\right)
 \sum_{m=1}^{\varrho\left( \mathbf{\mathcal{A}}_l\right)}\sum_{n=1}^{\tau_m \left( \mathbf{\mathcal{A}}_l\right)}
 \sum_{p=0}^{N-K} \frac{\left(\frac{\gamma_{{\tt th}}}{\beta_{llk}}\right)^p}{p!}\!\mathcal{X}_{m,n} \left(
 \mathbf{\mathcal{A}}_l\right)\frac{\mu_{l,m}^{-n}}{\left(n-1\right)!}
     \nonumber  \\
   & \hspace{5cm}\times
 \int_0^\infty \left(1/\Pu+z\right)^p z^{n-1}\exp\left(-z\left(\frac{1}{\mu_{l,m}}+\frac{\gamma_{{\tt th}}}{\beta_{llk}}\right)\right) dz.
\end{align}

Using the binomial expansion and \cite[Eq.~(3.326.2)]{GR:07:Book},
we arrive at the desired result \eqref{eq Pro Out 1}.

% \bibliographystyle{IEEEtran}
% \bibliography{IEEEabrv,CCTLABBiblio}

\clearpage
\newpage

%C. J. Chen, Performance Analysis of Scheduling in Multiuser MIMO
%Systems with Zero-forcing receivers

\clearpage

\clearpage

\clearpage

%\begin{figure}[t]
%    \centering
%    \centerline{\includegraphics[width=0.8\textwidth]{Figures/Rate_determ.eps}}
%    \caption{Analytical sum rate and deterministic approximation versus the number of BS antennas $N$.
%    ($L=4$, $\kappa=2$, $\mathsf{SNR} = 10$ dB, $a=0.1$, $0.3$, and $0.5$).}
%    \label{fig:4}
%\end{figure}

\clearpage

\clearpage

%\begin{figure}[t]
%    \centering
%    \centerline{\includegraphics[width=0.8\textwidth]{Figures/Outage_M.eps}}
%    \caption{Simulated and analytical outage probability versus the number of BS antennas $N$. ($L=4$, $K=10$, $\mathsf{SNR}= 0$ dB, $
%    \gamma_{\tt th} = 0$ dB, $a=0.1$, $0.3$, and $0.5$).}
%    \label{fig:7}
%\end{figure}

%\begin{figure}[t]
%    \centering
%    \centerline{\includegraphics[width=0.8\textwidth]{Figures/Outage_R.eps}}
%    \caption{Outage probability versus data rate. The circles represent $95\%$-likely rates. ($L=4$, $K=10$, $a=0.1$, and $\mathsf{SNR} = 0$ dB).}
%    \label{fig:8}
%\end{figure}

\clearpage

\clearpage

\begin{table}[h]
    \caption{
        Uplink performance of ZF with frequency-reuse factors $1, 3$, and $7$,
        for $\Pu=10 \mathrm{dB}$, $N=20, 100$, $\sigma_{\mathrm{shadow}} =8 \mathrm{dB}$, and $\nu=3.8$
    }
    \centerline{\includegraphics[width=1\textwidth]{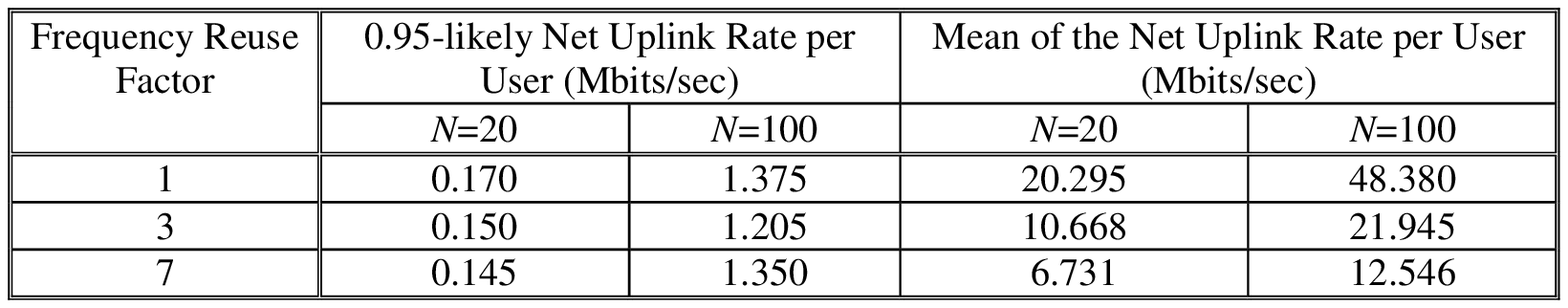}}
    \label{table:1}
\end{table}

\end{document}